\documentclass[a4paper,11pt]{article}
\pdfoutput=1
\usepackage{./auxi/jheppub}
\usepackage[utf8]{inputenc}

\usepackage{simplewick}
\usepackage{dsfont}
\usepackage{float}
\usepackage{pgfplots}
\usepackage{slashed}
\pgfplotsset{compat=1.14}
\usepackage{tikz}
\usetikzlibrary{shapes}

\let\oldbfseries=\bfseries
\let\oldmdseries=\mdseries
\let\oldnormalfont=\normalfont
\renewcommand{\bfseries}{\oldbfseries\boldmath}
\renewcommand{\mdseries}{\oldmdseries\unboldmath}
\renewcommand{\normalfont}{\oldnormalfont\unboldmath}

\makeatletter
\newlength{\apb@width}
\newcommand{\autoparbox}[2][c]{\settowidth{\apb@width}{#2}\parbox[#1]{\apb@width}{#2}}

\makeatother

\def \ph{\phantom}
\DeclareMathOperator{\ad}{ad}
\DeclareMathOperator{\tr}{tr}

\def\Cm{{\mathcal{C}}}
\def\Dm{{\mathcal{D}}}

\def\Fm{{\mathcal{F}}}
\def\Gm{{\mathcal{G}}}

\def\Jm{{\mathcal{J}}}
\def\Km{{\mathcal{K}}}

\def\Mm{{\mathcal{M}}}
\def\Nm{{\mathcal{N}}}
\def\Om{{\mathcal{O}}}
\def\Pm{{\mathcal{P}}}
\def\Qm{{\mathcal{Q}}}
\def\Rm{{\mathcal{R}}}
\def\Sm{{\mathcal{S}}}

\def\a{{\alpha}}
\def\b{{\beta}}
\def\c{\gamma}

\def\ad{{\dot{\alpha}}}

\def\ph{\phantom}

\newcommand{\beq}{\begin{equation}}
\newcommand{\eeq}{\end{equation}}

\definecolor{nicegreen}{rgb}{0.1,0.6,0.1}

\newmuskip\pFqskip
\mathchardef\pFcomma=\mathcode`,

\makeatletter
\renewcommand*\env@matrix[1][\arraystretch]{%
  \edef\arraystretch{#1}%
  \hskip -\arraycolsep
  \let\@ifnextchar\new@ifnextchar
  \array{*\c@MaxMatrixCols c}}
\makeatother

\newcommand{\diagramEnvelope}[1]{#1}

\tikzset{
  vertex/.style={
    circle,
    draw,
    fill=black,
    inner sep=1pt,
    minimum size=1pt
  },
  starrr/.style={
    circle,
    draw,
  },
  point/.style={
    inner sep=0pt,
    minimum size=0pt
  },
  prop/.style={
    thick
  },
  sus/.style={
    thick,
    dashed
  },
}

\newcommand{\half}{\frac{1}{2}}
\newcommand{\Pop}{\mathcal{P}}
\newcommand{\Kop}{\mathcal{K}}
\newcommand{\Dop}{\mathcal{D}}
\newcommand{\Mop}{\mathcal{M}}
\newcommand{\Rop}{\mathcal{R}}
\newcommand{\Qop}{\mathcal{Q}}
\newcommand{\Qbop}{\bar{\mathcal{Q}}}
\newcommand{\Sop}{\mathcal{S}}
\newcommand{\Sbop}{\bar{\mathcal{S}}}
\newcommand{\Oop}{\mathcal{O}}

\graphicspath{{./figures/}}

\title{\boldmath Superconformal Boundaries in $4-\epsilon$ dimensions}

\author{Aleix Gimenez-Grau,}
\author{Pedro Liendo,}
\author{Philine van Vliet}

\affiliation{DESY Hamburg, Theory Group, Notkestra{\ss}e 85, D-22607 Hamburg, Germany}

\emailAdd{aleix.gimenez@desy.de}
\emailAdd{pedro.liendo@desy.de}
\emailAdd{philine.vanvliet@desy.de}

\preprint{DESY 20-215}

\abstract{
Boundaries in three-dimensional $\mathcal{N}=2$ superconformal theories may preserve one half of the original bulk supersymmetry.
There are two possibilities which are characterized by the chirality of the leftover supercharges. Depending on the choice, the remaining $2d$ boundary algebra exhibits $\mathcal{N}=(0,2)$ or $\mathcal{N}=(1,1)$ supersymmetry. 		 
In this work we focus on correlation functions of chiral fields for both types of supersymmetric boundaries. We study a host of correlators using superspace techniques and calculate superconformal blocks for two- and three-point functions. 	  
For $\mathcal{N}=(1,1)$ supersymmetry, some of our results can be analytically continued in the spacetime dimension while keeping the codimension fixed.  This opens the door for a bootstrap analysis of the $\epsilon$-expansion in supersymmetric BCFTs. Armed with our analytically-continued superblocks, we prove that in the free theory limit two-point functions of chiral (and antichiral) fields are unique. The first order correction, which already describes interactions, is universal up to two free parameters. As a check of our analysis, we study the Wess-Zumino model with a supersymmetric boundary using Feynman diagrams, and find perfect agreement between the perturbative and bootstrap results.}

\begin{document} 
\setcounter{tocdepth}{2}
\maketitle
\setcounter{page}{1}

\section{Introduction}
\label{sec:intro}

Conformal field theories with boundaries have a wide variety of applications that range from condensed matter to string theory. In recent years the conformal bootstrap has emerged as a powerful tool to study CFTs and has also been applied to boundary conformal theories (BCFTs). The numerical bootstrap for BCFTs was originally implemented in \cite{Liendo:2012hy,Gliozzi:2015qsa}, while analytical approaches, which include the $\epsilon$-expansion bootstrap \cite{Liendo:2012hy,Bissi:2018mcq}, and the construction of exact linear functionals \cite{Kaviraj:2018tfd,Mazac:2018biw}, have also been explored. A closely related line of research is to study the dynamics of free bulk theories with non-trival dynamics localized in the boundary \cite{Prochazka:2019fah,Giombi:2019enr,Herzog:2017xha,Herzog:2018lqz,DiPietro:2019hqe,Behan:2020nsf}.

In this work we study supersymmetric boundaries for three-dimensional models with $\mathcal{N}=2$ supersymmetry, a setup that has received particular attention in the context of infrared dualities \cite{Gadde:2013wq,Okazaki:2013kaa,Aprile:2016gvn,Dimofte:2017tpi}, and localization \cite{Sugishita:2013jca,Yoshida:2014ssa}. There are two ways in which supersymmetry can be preserved when a boundary is introduced: one choice preserves supercharges of the same chirality which define a $2d$ $\Nm=(0,2)$ subalgebra, while the other choice is non-chiral and describes a  $2d$ $\Nm=(1,1)$ subalgebra. We will study the kinematical constraints on correlators for both choices, with a particular emphasis on two-point functions. As is well known, in the presence of a boundary two-point correlators are not fixed by symmetry, but depend on a conformal invariant. They contain non-trivial dynamics akin to four-point functions in homogeneous CFTs, which is captured by the existence of two inequivalent conformal block expansions. One possibility is to fuse the two local operators together and calculate the resulting one-point functions in the presence of the boundary. Another option is to expand a local operator as an infinite sum of boundary excitations, and calculate the resulting two-point functions on the boundary. Consistency between the two decompositions is the starting point of the bootstrap program for BCFT.

Our main focus will be chiral fields, which are short operators of the bulk superconformal algebra killed by half of the supercharges, and whose conformal dimension is fixed by the $R$-symmetry. As usual in the bootstrap, it is essential to calculate the relevant superconformal blocks. Bosonic blocks for BCFT two-point functions have been known for a long time \cite{McAvity:1995zd}, however less work has been done on supersymmetric models, the sole exception being boundaries in $\Nm=4$ SYM \cite{Liendo:2016ymz}. Attempts to formalize the study of superconformal blocks include analytic superspace \cite{Doobary:2015gia} and the connection to Calogero-Sutherland models \cite{Buric:2019rms,Buric:2020buk}. Here we start our analysis using standard superspace techniques, and calculate superblocks using the Casimir approach \cite{Dolan:2003hv}.
The superspace analysis will be uniform for the $\Nm = (0,2)$ and $\Nm = (1,1)$ subalgebras, but it turns out that the $\Nm = (1,1)$ blocks have the interesting property that they can be analytically continued across dimensions.
In more detail, there is a unique half-BPS boundary in $4d$ $\Nm = 1$ which is non-chiral, and that can be interpolated to the $\Nm = (1,1)$ boundary in $3d$.
This is the BCFT counterpart of the results obtained in \cite{Bobev:2015jxa}, where the bulk superconformal blocks were continued in $d$. Even though conformal symmetry is subtle in non-integer dimensions,\footnote{See \cite{Hogervorst:2015akt} for discussions on non-integer $d$ and \cite{Binder:2019zqc} for non-integer $N$ (in the context of $O(N)$ models).} conformal blocks are usually analytic in all their quantum numbers.\footnote{In the case of the defects both dimension and codimension appear as parameters in the blocks \cite{Billo:2016cpy,Isachenkov:2018pef}.}

Armed with the analytic continuation we tackle the $\epsilon$-expansion for models that satisfy our constraints. Using minimal assumptions, we prove that two-point functions of free chiral and antichiral fields are completely fixed. At leading order in $\epsilon$, which already corresponds to an interacting fixed point, we prove that the two-point functions are universal up to two free parameters: the anomalous dimension of the lowest-lying bulk field, and the anomalous dimension of the lowest-lying boundary field. The solution is non-trivial and contains an infinite number of conformal blocks, and therefore can be used to extract an infinite amount of CFT data.

As a check of our general order $\epsilon$ result, we concentrate on the Wess-Zumino model with cubic superpotential, which is a prime example of a critical system that preserves four supercharges. 
Using the results of~\cite{Bilal:2011gp}, we construct an explicit Lagrangian model with boundary degrees of freedom that exhibits all the symmetries of our setup. We use this model to perform a Feynman diagram calculation at one-loop order, and confirm that the perturbative result is in perfect agreement with our bootstrap prediction.

The outline of the paper is as follows.
In section~\ref{sec:preliminaries} we summarize the differences between the $\Nm=(0,2)$ and $\Nm = (1,1)$ boundaries and introduce the crossing equations for BCFT.
In section~\ref{sec:3d} we carry out a detailed study of correlation functions and superconformal blocks of these $3d$ models.
In section~\ref{sec:4-eps} we rederive the superconformal blocks with a new method that is applicable to any $3 \le d \le 4$, and use them to bootstrap two-point functions of chiral operators in the $\epsilon$ expansion.
Finally, in section~\ref{sec:wess-zumino} we compute the same two-point functions for the Wess-Zumino model using Feynman diagrams.
We conclude with some possible future directions in section~\ref{sec:conclusions} and we relegate some technical details to the appendices.

\section{Preliminaries}
\label{sec:preliminaries}

In this preliminary section we introduce the symmetry algebra in the bulk, and the two possible half-BPS subalgebras preserved by a supersymmetric boundary. We also introduce chiral fields and review the standard bootstrap equations for two-point functions in BCFT. 

\subsection{Superconformal boundaries in \texorpdfstring{$3d$}{3d}}
\label{subsec:3dbulk}

There are two inequivalent half-BPS boundaries that one can consider in $3d$ $\Nm=2$ superconformal theories, which are commonly denoted as $\Nm = (0,2)$ and $\Nm = (1,1)$ boundaries (see \cite{Sugishita:2013jca,Yoshida:2014ssa,Gadde:2013wq,Okazaki:2013kaa,Aprile:2016gvn,Dimofte:2017tpi,Brunner:2019qyf} for related work). 
The cleanest way to understand their differences is at the level of the commutation relations of their algebras.

Let us start by reminding the reader about the main features of  $3d$ $\Nm=2$ superconformal symmetry.
Besides the conformal generators $\Dm, \Pm_\mu, \Km_\mu, \Mm_{\mu\nu}$, the superconformal algebra has four Poincaré supercharges $\Qm_\a, \bar\Qm_\a$, and four superconformal partners $\Sm_\a, \bar\Sm_\a$.
There is an extra $U(1)$ symmetry generated by $\Rm$ under which $\Qm_a,\Sm_a$ have charge $-1$, and $\bar \Qm_a, \bar \Sm_a$ have charge $+1$.
The precise commutation relations with a summary of our conventions are presented in appendix~\ref{app:3d}.
The representation theory of $3d$ $\Nm=2$ is well known and can be found for example in~\cite{Cordova:2016emh}.

When we restrict ourselves to three dimensions, we take the superconformal boundary to be located at $x_2 \equiv x_\bot = 0$.
It is clear that the bosonic subalgebra is generated by $\Pop_a$, $\Kop_a$, $\Mop_{ab}$, and $\Dop$, where $a = 0, 1$ runs over directions parallel to the boundary.
We are now ready to introduce the two inequivalent boundary-preserving superalgebras, which differ only by the choice of fermionic generators.

\paragraph{The \texorpdfstring{$\Nm=(0,2)$}{N=(0,2)} boundary:}

The first possibility is to choose the following fermionic generators: $\Qop_{2}$, $\Qbop_2$, $\Sop^{2}$, $\Sbop^2$.
The precise commutation relations can be obtained by restricting the full superconformal algebra presented in appendix~\ref{sec:app-algebra}. 
The following ones are of particular importance:
\begin{align}
\begin{split}
 & \{ \Qm_2, \bar \Qm_2 \} = -2 (\Pm_0 + \Pm_1), \\
 & \{ \Qm_2, \bar \Sm^2 \} = -2i \Dm + 2\Rm + 2i \Mm_{01}, \\
 & \{ \bar \Qm_2, \Sm^2 \} = -2i \Dm - 2\Rm + 2i \Mm_{01}.\label{eq:02alg}
\end{split}
\end{align}
From the first equation, we notice that $\Pm_2$ does not appear on the right-hand side. This was to be expected, since translations in the $x_2 = x_\bot$ direction are not preserved.
Regarding the second and third equations, it is crucial to notice the appearence of $\Rm$.
Physically, it means that the $\Nm=(0,2)$ boundary preserves $R$-symmetry, a property that strongly constrains which correlation functions are non-vanishing.
For example, we will often be concerned with bulk operators $\Om_{r}$ with charge $r$.
From the above discussion, it follows that one-point functions $\langle \Om_r \rangle = 0$ unless $r = 0$, and similarly two-point functions $\langle \Om_{r_1} \Om_{r_2} \rangle = 0$ unless $r_1 + r_2 = 0$.

\paragraph{The \texorpdfstring{$\Nm=(1,1)$}{N=(1,1)} boundary:}

The second possibility is to choose the following fermionic generators:
\begin{align}
\begin{split}
 	\tilde{\Qop}_1 
 	\equiv \frac{1}{\sqrt{2}} \left(\Qop_1 + \bar{\Qop}_1 \right), 
 	&\quad \tilde{\Qop}_2 \equiv \frac{i}{\sqrt{2}} \left(\Qop_2 - \bar{\Qop}_2 \right),\\
	\tilde{\Sop}^1 
	\equiv \frac{1}{\sqrt{2}} \left(\Sop^1 + \bar{\Sop}^1\right), 
	&\quad \tilde{\Sop}^2 
	\equiv \frac{i}{\sqrt{2}} \left(\Sop^2 - \bar{\Sop}^2 \right). \label{eq:11alg}
\end{split}
\end{align}
Once again, the full set of commutation relations can be obtained from the formulas in appendix~\ref{sec:app-algebra}.
The non-vanishing anticommutators are
\begin{align}
\begin{split}
 	\{ \tilde{\Qop}_{\alpha},\tilde{\Qop}_{\beta}\} &= 2 (\gamma^a)_{\alpha \beta} \mathcal{P}_a\,,\\
	\{ \tilde{\Qop}_{1},\tilde{\Sop}^{1} \} &= - 2i (\Dop + \Mop_{01})\,, \\
	\{ \tilde{\Qop}_{2},\tilde{\Sop}^{2} \} &= - 2i (\Dop - \Mop_{01})\,.
\end{split}
\end{align}
As before, $\Pm_2$ is not part of the algebra since $a = 0,1$. 
Interestingly, the second anticommutator does not contain $\Rm$, since $R$-symmetry is broken by the $\Nm = (1,1)$ boundary.
In this case, charged bulk operators can have one- and two-point functions that would be forbidden by charge conservation, namely  $\langle \Om_r \rangle \ne 0 \ne \langle \Om_{r_1} \Om_{r_2} \rangle$ for any values of the charges.

\subsection{Chiral primaries in superconformal theories}

As announced before, we will mostly focus on chiral primary operators $\phi$ and their complex conjugates.
Often, we will call these operators ``chirals'' and ``antichirals'' for simplicity.
These are short multiplets of the superconformal algebra killed by half of the supercharges: 
\begin{align}
 [\bar \Qm_\a, \phi(0)] = 0\,, \qquad
 [\Qm_\a, \bar \phi(0)] = 0\,,
\end{align}
and whose conformal dimension and $R$-charge are related to each other. 
For general spacetime dimension one obtains
\begin{align}
\label{eq:dim-chiral-across}
   \Delta_{\phi} 
 = \frac{d-1}{2} r_\phi\,, \qquad
 \Delta_{\bar\phi} 
 = - \frac{d-1}{2} r_{\bar\phi}\,.
\end{align}
There is a consistent way to define chiral multiplets in any, in principle continuous, number of dimensions, a fact that will play a significant role in section~\ref{sec:4-eps}.

Chiral operators are ubiquitous in the study of SCFTs and they are present in most known models.
A textbook example of a $4d$ Lagrangian with $\Nm = 1$ supersymmetry is to consider chiral fields in superspace with some non-linear interaction. 
We will consider a simple example of this in section~\ref{sec:wess-zumino}, where we study the Wess-Zumino model, i.e.\ a single chiral multiplet with a cubic superpotential. This model flows to an interacting fixed point, which can be described perturbatively in the $\epsilon$-expansion using weakly-coupled chiral fields. It turns out that the $\epsilon$-expansion can be generalized to include boundaries, a fact that we will explore using the bootstrap results obtained in this work.

An important property of chiral operators is that they often satisfy non-trivial chiral-ring relations.
These relations are dynamic and imply that certain chiral operators might disappear from an OPE, for example $\phi_3 \notin \phi_1 \times \phi_2$, even if this is not forbidden by superconformal symmetry.
The Wess-Zumino model in $3d$ is a simple SCFT with chiral-ring relations. The chiral ring of this model is generated by $\phi$ together with the relation $\phi^2 \notin \phi \times \phi$.
In the numerical bootstrap analysis of~\cite{Bobev:2015vsa,Bobev:2015jxa}, the chiral-ring relation provided strong evidence that a kink in the numerical plots described the Wess-Zumino model. In section~\ref{sec:fin-res} we will notice that our perturbative results are also consistent with the same chiral-ring relation. 
More complicated examples of chiral-ring relations can be found in~\cite{Baggio:2017mas} where the authors studied numerically a $3d$ conformal manifold parametrized by the complex gauge coupling $\tau$.
Chiral-ring relations of bulk operators could also be used to extract information of a theory living on the boundary, similar in spirit to the work of~\cite{Lauria:2020emq,Behan:2020nsf}.\footnote{We thank Edo Lauria for discussions on this idea.}

In this work we will not explore all these questions yet, but they motivated us to study this setup.
Here we will work out basic kinematical constraints and use the bootstrap to study the dynamics of a single chiral field. Possible future directions and applications of our results will be discussed in the conclusions. Before we jump to the main analysis, let us first review the bootstrap approach for BCFT, which will be one of our main tools.

\subsection{Crossing symmetry in BCFT}
In this section we will review crossing symmetry for generic, non-supersymmetric boundary CFTs.
There are two relevant symmetry algebras to study BCFT.
The first one contains the $d$-dimensional conformal group, and it describes physics far away from the boundary.
In particular, bulk local operators $\Om(x)$ transform in irreducible representations of this algebra, and are labeled by a conformal dimension $\Delta$ and spin $\ell$.
There can also be physical excitations localized on the boundary, which are represented by local operators $\hat\Om(x_a, x^\bot\!=\!0)$.
These boundary operators transform as irreducible representations of the symmetry algebra that preserves the boundary, namely they have conformal dimension $\hat \Delta$ and $d-1$ dimensional spin $j$.

Correlation functions can be constructed with arbitrary combinations of bulk and boundary operators.
As usual, conformal symmetry puts strong constraints on the form of these correlation functions.
For example, the one-point function and the bulk-to-boundary correlator of a bulk scalar are fixed up to a constant~\cite{McAvity:1995zd} 
\begin{align}
	\label{eq:1pt-bos}
	\langle \Om(x) \rangle 
	= \frac{a_\Om}{(2 x^\bot)^{\Delta}} \, , \qquad
	\langle \Om(x_1) \hat \Om(x_2^a) \rangle 
	= \frac{b_{\Om\hat\Om}}
	{(2 x_1^\bot)^{\Delta-\hat\Delta} 
		\left( (x_{12}^a)^2 + (x_1^\bot)^2 \right)^{\hat\Delta}} \, .
\end{align}
For more general correlation functions the situation is more involved, because they can depend on conformal invariants.
For example, a two-point function of bulk scalars depends on an arbitrary function of the invariant $\xi$:
\begin{align}
\label{eq:two-pt-bos}
	\langle \Om_1(x_1) \Om_2(x_2) \rangle
	= \frac{\Fm(\xi)}{(2x_1^\bot)^{\Delta_1} (2x_2^\bot)^{\Delta_2}} \, , 
	\qquad
	\xi = \frac{(x_1 - x_2)^2}{4 x_1^\bot x_2^\bot} \, .
\end{align}
Knowledge of $\Fm(\xi)$ is equivalent to knowing the full two-point correlator.
The function $\Fm(\xi)$ is far from arbitrary; it is heavily constrained by crossing symmetry and it is the main subject of study in the bootstrap program for BCFT.

The main ingredient to derive the crossing equation is the operator product expansion (OPE).
It is well known that one can rewrite a product of two bulk local operators as an infinite sum of individual bulk local operators using the standard OPE. In the presence of a boundary there is a second possible expansion, the boundary operator expansion (BOE), in which one bulk 
local operator is replaced by a sum of operators that are localized in the boundary. 
In terms of equations, these two OPEs are
\begin{align}\label{eq:bulkbdyOPE}
	\begin{split}
		\mathcal{O}_1(x) \mathcal{O}_2(0) 
		& = \frac{1}{x^{2\Delta}} 
		+ \sum_{\Om} \lambda_{\Om_1\Om_2\Om} 
		C [x, \partial_x] \Om(0) \, , \\
		\mathcal{O}(x) 
		& = \frac{a_\Om}{(2 x_\bot)^\Delta} 
		+ \sum_{\hat\Om} b_{\Om\hat\Om} D[x_\bot , \partial_a ] \hat{\mathcal{O}} (x^a) \, .
	\end{split}
\end{align}
The sums run only over conformal primaries, and the contributions of the descendants are captured by the differential operators $C$ and $D$ which are completely fixed by conformal symmetry.

The power of the OPE is that it allows us to evaluate higher-point functions using lower-point correlators, provided we know the spectrum of the theory and all the OPE coefficients $a$, $b$ and $\lambda$.
In the example of a bulk two-point function, there are two different decompositions possible:
\begin{equation}
\label{eq:bos-crossing-eq}
\mathcal{F} (\xi) 
= \sum_\Om a_\Om \lambda_{\Om_1\Om_2\Om}  f_{\Delta}(\xi)
= \sum_{\hat\Om} b_{\Om_1\hat\Om} b_{\Om_2\hat\Om} \hat f_{\hat\Delta}(\xi) \, .
\end{equation}
The objects $f_{\Delta}(\xi)$ and $\hat f_{\hat\Delta}(\xi)$ are called conformal blocks, which we review in  appendix~\ref{sec:bosblocks}.
Equation \eqref{eq:bos-crossing-eq} is called the ``crossing equation'', and it provides non-trivial constraints on the spectrum and CFT data of boundary conformal field theories.

The above discussion was completely general, and it applies to any conformal field theory with a conformal boundary.
The main goal of the present paper is to specialize it to superconformal boundaries, in which case the crossing equation~\eqref{eq:bos-crossing-eq} can be constrained even further.
The reason is that supersymmetry relates the OPE coefficients of different conformal primaries that belong to the same supermultiplet, which means that we can organize the expansion in terms of superconformal blocks $F_\Delta(\xi)$ and $\hat F_{\hat\Delta}(\xi)$.
These new objects are linear combinations of the bosonic blocks $f_\Delta(\xi)$ and $\hat f_{\hat\Delta}(\xi)$ with coefficients fixed by supersymmetry.
In sections~\ref{sec:3d} and~\ref{sec:4-eps} we will compute these objects in $d=3$ and in $3 \le d \le 4$ respectively, which will allow us to study the bootstrap equations analytically in section~\ref{sec:eps-bootstrap}.

\section{Boundaries in three dimensions}
\label{sec:3d}

\subsection{Superspace analysis}

Let us start by studying correlators for both types of boundary conditions using superspace techniques. We introduce a standard Minkowski superspace in which each supercharge  $\Qop_\alpha, \Qbop_\alpha$, where $\alpha = (1,2) = (-,+)$, has a Grassmann variable $\theta^\a$, $\bar\theta^\a$ associated to it. This setup is enough for our purposes, because we will mostly study correlators of scalar operators in a system with minimal supersymmetry.\footnote{See~\cite{Lauria:2018klo,Herzog:2020bqw} for studies of non-supersymmetric two-point functions of arbitrary spin.} 
Our superspace then consists of three spacetime coordinates $x^{\mu}$ and four Grassmann coordinates $\theta^{\alpha}$ and $\bar{\theta}^{\alpha}$ which we collect as follows:
\begin{equation}
z=(x^\mu, \theta^{\alpha}, \bar{\theta}^{\alpha})\, ,
\end{equation}
where $\mu = 0,1,2$.
We can convert spinor indices $\alpha,\beta$ into vector indices $\mu,\nu$ by means of the gamma matrices $(\gamma^{\mu})_{\alpha \beta}$.
The form of these matrices, together with further conventions regarding raising, lowering and contracting indices, can be found in appendix~\ref{sec:conventions}. 

The differential form of the (super)translations acting on fields $\mathcal{O}(z)$ is standard
\begin{align}
	\left[\Pop_\mu, \Oop(z)\right] &=  i \partial_\mu \Oop(z)\,, \\
	\left[\Qop_\alpha,\Oop(z)\right] &= \left(\partial_\alpha + i (\gamma^{\mu})_{\alpha \beta} \bar{\theta}^{\beta} \partial_\mu \right)\Oop(z)\,, \\ \left[\Qbop_\alpha,\Oop(z)\right] &= - \left(\bar{\partial}_\alpha + i (\gamma^{\mu})_{\alpha \beta} \theta^{\beta} \partial_\mu\right)\Oop(z)\,.
\end{align}
From the bulk algebra it is easy to derive the form of all the other differential operators, which we list in appendix~\ref{sec:diff-ops}.
The action of the covariant derivatives is also standard
\begin{equation}
D_{\alpha} \Oop(z)
= \left(\partial_{\alpha}  - i (\gamma^{\mu})_{\alpha \beta} \bar{\theta}^{\beta} \partial_{\mu}\right)\Oop(z)\,, \quad 
\bar{D}_{\alpha} \Oop(z) 
= - \left(\bar{\partial}_{\alpha}  - i (\gamma^{\mu})_{\alpha \beta} \theta^{\beta} \partial_{\mu}\right)\Oop(z)\,,
\end{equation}
and as usual, they anticommute with the action of supertranslations.
The main focus of this paper is on chiral and antichiral operators (see section~\ref{sec:preliminaries}), which are defined in superspace as
\begin{align}
\label{eq:chirality}
 \bar D_\a \Phi (z) = 0\,, \qquad
 D_\a \bar \Phi (z) = 0\,.
\end{align}
In order to work with chiral operators it is useful to work with chiral/antichiral coordinates defined as
\begin{align}
\label{eq:chiraldist}
 y^\mu = x^\mu - i \gamma^\mu_{\a\b} \theta^\a \bar \theta^\b, \qquad
 \bar y^\mu = x^\mu + i \gamma^\mu_{\a\b} \theta^\a \bar \theta^\b.
\end{align}
In terms of these coordinates, a chiral field depends only on $\Phi(y,\theta)$ and similarly for the antichiral field $\bar\Phi(\bar y, \bar\theta)$.
If we consider two points, we can also define supersymmetric invariant distances with well-defined chirality:\footnote{Note that $y_{12}^\mu \ne y^\mu_1 - y^\mu_2$, we hope the notation will not create confusion.}
\begin{align}
\label{eq:chiraldistmu}
y_{12}^{\mu} &= x_{12}^{\mu} - i (\gamma^{\mu})_{\alpha \beta}\left( \theta_{1}^\alpha \bar{\theta}_{1}^\beta + \theta_{2}^\alpha \bar{\theta}_{2}^\beta - 2 \theta_{1}^\alpha \bar{\theta}_{2}^\beta\right)\,, \\
\bar{y}_{12}^{\mu} &= x_{12}^{\mu} + i (\gamma^{\mu})_{\alpha \beta}\left( \theta_{1}^\alpha \bar{\theta}_{1}^\beta + \theta_{2}^\alpha \bar{\theta}_{2}^\beta + 2 \bar{\theta}_{1}^\alpha \theta_{2}^\beta\right)\,.
\end{align}
These distances are chiral at one point and antichiral at the other, namely
\begin{equation}
 \bar D_\alpha^{(1)} y_{12}^\mu = D_\alpha^{(2)} y_{12}^\mu = 0\,, \qquad
 D_\alpha^{(1)} \bar y_{12}^\mu = \bar D_\alpha^{(2)} \bar y_{12}^\mu = 0\,.
\end{equation}

Introducing a boundary will generally break supersymmetry in the bulk. In this paper we study a special class of boundaries that preserve one half of the supersymmetry. As already discussed, they are characterized by $2d$ algebras with $\mathcal{N} = (0,2)$ and $\mathcal{N} = (1,1)$ supersymmetry respectively. The two boundaries have distinct features that we discuss in detail below, the most prominent being that the $\mathcal{N} = (1,1)$ boundary breaks $R$-symmetry, while it is kept intact in the $\mathcal{N} = (0,2)$ case.
%
%
\subsection{The \texorpdfstring{$\mathcal{N} = (0,2)$}{N=(0,2)} boundary}
The $\mathcal{N} = (0,2)$ boundary preserves the supercharges $\Qop_{+}, \Qbop_{+}$, resulting in the algebra given in~\eqref{eq:02alg}. The bulk superspace can be split into coordinates parallel and perpendicular to the boundary. The parallel coordinates are
\begin{equation}
\left(\theta^{+}, \bar{\theta}^{+}, x^{a}\right), \quad a = 0,1,
\end{equation}
while the perpendicular coordinates read
\begin{equation}
\big(\theta^{-}, \bar{\theta}^{-}, x^\bot \equiv x^{2}\big).
\end{equation}
As was the case for the bulk theory, it is convenient to define supersymmetric, chiral, and antichiral perpendicular distances. The supersymmetric distance is
\begin{equation}\label{eq:ppdist02}
z^\bot = x^\bot - i \theta^{\alpha} \bar{\theta}_\alpha
\end{equation}
and the chiral $y^\bot \equiv y^2$ and antichiral $\bar{y}^\bot \equiv \bar y^2$ perpendicular distances can be read off from~\eqref{eq:chiraldist}.
Note that $z^\bot$ is invariant under the boundary (super)translations $\Pop_a, \Qop_{+}$ and $\Qbop_{+}$, while $y^\bot$ and $\bar{y}^\bot$ are not. 
The component expansion of a chiral field $\Phi$ takes the familiar form
\begin{equation}
\Phi(y,\theta) = \phi(y) + \theta^{+} \psi_{+}(y) + \theta^{-} \psi_{-}(y) + \theta^{+} \theta^{-} F(y)\,,
\end{equation}
where $\phi$ is a complex boson, $\psi_\alpha$ a complex fermion, and $F$ a complex auxiliary field.
It will be convenient to decompose this bulk chiral supermultiplet $\Phi$ in terms boundary supermultiplets, that transform irreducibly under the $(0,2)$ subalgebra~\cite{Brunner:2019qyf,Dimofte:2017tpi}
\begin{equation}
\label{eq:3dto2d_chiral}
\Phi = \hat{\Phi} +  \theta^{-} \hat{\Psi} + \ldots,
\end{equation}
where $\hat{\Phi}$ is a boundary chiral field, $\hat{\Psi}$ a boundary Fermi field, and the $\ldots$ stand for derivatives of $\hat\Phi$ parallel to the boundary. A similar expansion can be written for the antichiral bulk supermultiplet $\bar{\Phi}$. From now on, we will denote boundary multiplets and boundary fields with a hat.
One can straightforwardly derive a similar expansion for $\hat{\Phi}$ and $\hat{\Psi}$:
\begin{align}
\hat{\Phi} 
= \Phi
\Big|_{\theta^{-} = \bar{\theta}^{-} = 0} &= \phi + \theta^{+} \psi_{+} 
 + \ldots,\\
\hat{\Psi} = D_{-} \Phi \Big|_{\theta^{-} = \bar{\theta}^{-} = 0} &= \psi_{-} + \theta^{+} F 
+ \ldots,
\end{align}
where $F \sim \partial_{\bot} \phi$ on-shell and the $\ldots$ stand for terms with derivatives. The usual Neumann and Dirichlet boundary conditions can be neatly represented in terms of these superfields:
\begin{align}
\text{Neumann: } \quad
\partial_\bot \phi \Big|_{\partial} = 0, \quad 
\psi_{-} \Big|_{\partial} = 0 & \quad \to \quad
\hat{\Psi}|_{\partial} = 0, \\
\text{Dirichlet: } \quad \quad
\phi \Big|_{\partial} = 0, \quad
\psi_{+} \Big|_{\partial} = 0 & \quad \to \quad
\hat{\Phi}|_{\partial} = 0.
\end{align}
%
%
\subsubsection{One-point functions}
As reviewed in section \ref{sec:preliminaries}, scalar bulk operators can acquire a one-point function in the presence of a boundary. In the superspace setup we are considering, we expect on general grounds one-point functions of the form
\begin{equation}\label{eq:onepoint02}
\langle \Oop (z) \rangle = \frac{a_\Oop}{(z^\bot)^\Delta}\,,
\end{equation}
where $z^\bot$ is given in~\eqref{eq:ppdist02}.
For chiral fields, the chirality condition \eqref{eq:chirality} and conservation of $R$-symmetry imply that the one-point function vanishes: $a_\Phi = 0$.
%
\subsubsection{Bulk-to-boundary correlator}\label{sec:bulkbdytwopt02}
Similarly to the one-point function, we expect bulk-to-boundary correlators to be of the form
\begin{equation}\label{eq:bulkbdygen02}
\langle\Oop (z_1) \hat{\Oop} (z_2) \rangle = \frac{1}{(z_1^{\bot})^{\Delta - \hat{\Delta}} |y_{12}^2 \, \bar{y}_{12}^2|^{\hat{\Delta}/2}}\,g(\Theta_i)\,, 
\end{equation}
where $g$ is a function of possible nilpotent invariants  $\Theta_i$.

Again, the chirality condition \eqref{eq:chirality}  is extremely powerful and severely constrains the possible defect operators that can appear in the boundary OPE of a chiral field. From the expansion in \eqref{eq:3dto2d_chiral} we expect two types of boundary multiplets, and indeed there are two possible correlators consistent with all the symmetry constraints. One choice involves a scalar boundary multiplet\footnote{Whenever possible we supertranslate point 2 to the origin to simplify our formulas, but if necessary one can easily supertranslate back to a general frame.}
\begin{equation}\label{eq:bulkbdychir02}
\langle \Phi(y, \theta) \hat{\bar{\Phi}}_{\mathit{r}} (0) \rangle = \frac{b_{\Phi \hat{\bar{\Phi}}}}{ |y^\mu|^{2\Delta}}, 
\end{equation}
where $|y^\mu|$ is the norm of the chiral distance~\eqref{eq:chiraldist}, and the conformal dimensions are constrained by conservation of $R$-symmetry $ \hat{\Delta} = \Delta_\phi = \mathit{r}_\phi = r_{\hat\phi}$.

The other bulk-to-boundary two-point function involves the Fermi multiplet $\hat{\bar{\Psi}}$ whose highest weight carries spin:
\begin{equation}\label{eq:bulkbdyfermi02}
\langle\Phi (y, \theta) \hat{\bar{\Psi}} (0) \rangle = \frac{b_{\Phi \hat{\bar{\Psi}}} \gamma_{1 \beta}^\mu y_{\mu} \theta^\beta }{(y^{\bot})^{\Delta - \hat{\Delta} + \half} |y^\mu|^{2(\hat{\Delta} + \half)}}\,.
\end{equation}
Charge conservation implies $r_{\psi} = 1 - r_\phi$ but $\hat{\Delta}$ is not constrained to take a specific value, which means these multiplets are responsible for most of the operators that appear in the boundary block expansion of the two-point function of chiral fields. The power $\hat{\Delta} + \half$ indicates that the contributing field is not the primary, but a descendant (see equation \eqref{eq:Fermi_block} below).
%
%
\subsubsection{Two-point functions}
As reviewed in section \ref{sec:preliminaries}, bosonic two-point functions depend on a conformal invariant and therefore contain a large amount of dynamical information through their conformal block decompositions. As evident from our analysis so far, correlators of chiral fields are severely constrained by superconformal symmetry and their chirality condition. There is actually only one possible two-point invariant that satisfies all the superspace constraints:
\begin{equation}\label{eq:twopt-inv-02}
\xi = \frac{y_{12}^2}{4 y_{1}^\bot \bar{y}_{2}^\bot} \left( 1 + 2 i \frac{(y_{12}^0 + y_{12}^{1} )\theta_{1}^{-} \bar{\theta}_{2}^{-}}{y_{1}^\bot \bar{y}_{2}^\bot} + 2 i \frac{\theta_{1}^{+} \bar{\theta}_{2}^{-}}{\bar{y}_{2}^\bot} \right)\,.
\end{equation}
This is the unique ``supersymmetrization'' of the standard bosonic invariant. The most general two-point function of a chiral and an antichiral field then reads
\begin{equation}\label{eq:twopoint02}
\langle \Phi(y_1,\theta_1) \bar\Phi(0,\bar{y}_{2}^\bot, \bar\theta_2^-) \rangle 
= \left( \frac{\xi}{y_{12}^2} \right)^{\Delta} \mathcal{F}(\xi)\,,
\end{equation}
where $\mathcal{F}$ is an abitrary function of the superconformal invariant $\xi$. 
In equations~\eqref{eq:twopt-inv-02} and~\eqref{eq:twopoint02} we work in a frame where $\bar{y}_{2}^a = \bar\theta_2^+ = 0$, $(a = 0,1)$, but we keep the dependence on $\bar{y}_{2}^\bot$ and $\bar\theta_2^-$, since they are perpendicular coordinates and cannot be set to zero. 
Using a supertranslation one can find the two-point function in a frame with completely general $z_1$ and $z_2$, as will be needed below.
Two-point functions of two chiral (or two antichiral) fields are zero due to $R$-symmetry.
For more general external operators, for example long multiplets of the superconformal algebra, we expect a more complicated correlator involving nilpotent invariants, which then translates into superconformal blocks that have free parameters (see for example \cite{Cornagliotto:2017dup}). We will not consider more general correlators in this work, however our superspace setup could be used to study them in the future.
%
\subsubsection{Superconformal blocks}
\label{sec:susyblocks02}

We are now ready to obtain one of the main results of this section: the superconformal blocks associated to the two-point correlator $\mathcal{F}(\xi)$.
As reviewed in section~\ref{sec:preliminaries}, there are two conformal block expansions associated to the bulk and defect channel respectively. Bulk conformal blocks are eigenfunctions of the two-point bulk Casimir operator, while defect blocks are eigenfunctions of the defect Casimir.

\paragraph{Bulk channel:}
Let us start with the bulk channel, 
\begin{equation}\label{eq:bulkcaseqsusy}
\mathcal{C}_{\text{susy}}^{(12)}  \langle \Phi(z_1) \bar{\Phi}(z_2) \rangle 
= C_{\Delta, \ell, r}  \langle \Phi(z_1) \bar{\Phi}(z_2) \rangle\,,
\end{equation}
where the supersymmetric bulk Casimir is given by
\begin{align}\label{eq:bulkcassusy}
\mathcal{C}_{\text{susy}}^{(12)} &= - \Dop^2 -  \half \{ \Kop^{\mu}, \Pop_{\mu}\} +  \half \Mop^{\mu \nu} \Mop_{\mu \nu}  - \half \Rop^2 + \frac{1}{4} [\Sop^{\alpha}, \Qbop_{\alpha} ] + \frac{1}{4} [\Sbop^{\alpha}, \Qop_{\alpha} ]\, .
\end{align}
The superscript $(12)$ indicates that the operator acts on points $z_1$ and $z_2$. To avoid cluttering we wrote the superscript only on the Casimir, and omit it from the operators on the RHS. 
The eigenvalue reads
\begin{equation}\label{eq:bulkcaseigensusy}
C_{\Delta,\ell, \mathit{r}} 
= \Delta(\Delta - 1) + \ell (\ell + 1) - \frac{r^2}{2}\,.
\end{equation}
Evaluating ~\eqref{eq:bulkcaseqsusy} leads to a differential equation for the corresponding block $F_\Delta (\xi)$.
Our analysis implies the absence of nilpotent invariants when chiral fields are involved. This means that full superspace correlators can be reconstructed from those of the superprimaries and implies that a multiplet contributes only if its superprimary contributes. Because only scalars can acquire a one-point function in BCFT, we can safely set $\ell=0$ when looking for solutions to the Casimir equation. A standard approach to solve these equations is to recognize that superconformal blocks can be written as linear combinations of bosonic blocks. The superdescendants of a field $\Oop(z)$ can be generated by acting on the superprimary $\Oop(z)$ with the supercharges $\Qop, \Qbop$. This creates superdescendants of the schematic form $\Qop_1 \dots \Qbop_n \Oop(z)$.\footnote{In order to obtain proper conformal primaries (killed by $\Km$) the action of the $\Qop_i$ has to be corrected by terms containing the momentum generator $\Pm$.} We therefore make the following ansatz
\begin{equation}\label{eq:susyblockansatz}
F_{\Delta} (\xi) = f_{\Delta} (\xi) + c_0 f_{\Delta + \half} + c_1 f_{\Delta + 1}(\xi) + c_2 f_{\Delta + \frac{3}{2}}(\xi) + c_3 f_{\Delta + 2} (\xi) \, ,
\end{equation}
where $f_\Delta (\xi)$ are the bosonic blocks given in~\eqref{eq:bosbulkblocks}, and we fix the relative coefficients using \eqref{eq:bulkcaseqsusy}. 
The solution is easy to find
\begin{equation}\label{eq:bulkblocks02extra}
F_{\Delta} (\xi) = f_{\Delta} (\xi) -\frac{(\Delta -1) \Delta }{(2 \Delta -1) (2 \Delta +1)} f_{\Delta + 2} (\xi)\,,
\end{equation}
which corresponds to a long operator being exchanged in the $\phi \times \bar{\phi}$  OPE. There are also contributions from short multiplets, but they can be obtained from~\eqref{eq:bulkblocks02extra} evaluating $\Delta$ at the unitarity bound.
The selection rules of this OPE have been studied in the context of bulk four-point functions~\cite{Bobev:2015jxa} and our results are in perfect agreement with the literature. The block in equation~\eqref{eq:bulkblocks02extra} can be written as a single hypergeometric funtion
\begin{equation}
F_{\Delta} (\xi) = \xi^{\Delta /2}  \,  _2F_1 \Big(1 + \frac{\Delta }{2},\frac{\Delta }{2};\Delta +\frac{1}{2}; - \xi \Big).
\end{equation}
We will see that all of the two-point blocks derived in this section have this feature.

\paragraph{Boundary channel:} In the boundary channel the blocks are eigenfunctions of the boundary Casimir
\begin{equation}\label{eq:bdycassusy02}
\hat{\mathcal{C}}_{\text{susy}} = - \mathcal{D}^{2} -  \half \left\{ \mathcal{K}^{a} ,\mathcal{P}_{a} \right\}  + \half \mathcal{M}^{ a b} \mathcal{M}_{ab }  - \half  \Rop^2 + \frac{1}{4} \left( [\Sbop^{+}, \Qop_{+}] +  [\Sop^{+}, \Qbop_{+}] \right)\,,
\end{equation}
where now the operator acts at a single point:
\begin{equation}\label{eq:bdycaseqsusy02}
\hat{\mathcal{C}}_{\text{susy}}^{(1)}  \langle \Phi(z_1) \bar\Phi(z_2) \rangle 
= \hat{C}_{\hat{\Delta}, j,\mathit{r}}  
  \langle \Phi(z_1) \bar{\Phi}(z_2) \rangle\,.
\end{equation}
The eigenvalue depends on the conformal dimension $\hat\Delta$ of the exchanged boundary operator, as well as its parallel spin $j$ and its $R$-charge:
\begin{equation}\label{eq:bdycaseigensusy02}
\hat{C}_{\hat{\Delta},j,\mathit{r}} = \hat{\Delta} (\hat{\Delta} - 1) + j (j-1)  - \frac{\mathit{r}^2}{2}\,.
\end{equation}
Proceeding as before we make an ansatz for $\hat{F}_{\hat{\Delta}}$ in terms of bosonic blocks and fix the relative coefficients using \eqref{eq:bdycaseqsusy02}. Note that we only have to include conformal blocks up to dimension $\hat{\Delta} + 1$ in our ansatz, since the boundary only preserves half of the supercharges.
From section \ref{sec:bulkbdytwopt02} we know there are two types of boundary multiplets that can appear in the boundary expansion of a chiral field: a scalar $\hat{\Phi}$ and the Fermi multiplet $\hat{\Psi}$. We therefore expect two classes of solutions to the Casimir equation. Indeed, the solution corresponding to a chiral primary with $\mathit{r} = \mathit{r}_\phi$, $j = 0$ and $\hat{\Delta} = \Delta_\phi = \mathit{r}_{\phi}$, is given by 
\begin{equation}
\hat{F}^{\hat{\Phi}}_{\hat{\Delta}} (\xi) = \hat{f}_{\Delta_\phi} (\xi)\,.
\end{equation}
The second solution, with  $\mathit{r} = \mathit{r}_\phi - 1$, $j = \half$ corresponds to the Fermi field
\begin{equation}
\label{eq:Fermi_block}
\hat{F}^{\hat{\Psi}}_{\hat{\Delta}} (\xi) = \hat{f}_{\hat{\Delta}+ \half} (\xi)\,.
\end{equation}
Notice that the $ \half$ in the argument indicates that the highest weight does not contribute, but a descendant (as expected).
%
\subsubsection{Three-point functions}\label{sec:threept02}
Although not our main topic, let us also analyze three-point correlators involving one bulk field and two boundary fields. An interesting application for these correlators is to impose that the bulk field is free, and to study the corresponding constraints on the boundary three-point couplings \cite{Lauria:2020emq,Behan:2020nsf}.
For the rest of this section we will choose a frame where  $x^{a}_2 = \theta^{+}_2 = \bar{\theta}^{+}_2 = \theta^{+}_3 = \bar{\theta}^{+}_3 = 0, x^{a}_3 \to \infty$. By imposing that the bulk field is chiral we obtain 
\begin{equation} \label{eq:three-pt-PO2O3}
\langle \Phi(y, \theta) \hat{\Oop}_{2,j} (0, \omega) \hat{\Oop}_{3} (\infty) \rangle =  \frac{\left(y^a \omega_a\right)^j} {(y^{\bot})^{\Delta + \hat{\Delta}_{23}} |y^a|^j} \mathcal{F}^{\text{3pt}} (\chi)\,.
\end{equation}
The second operator has arbitrary parallel spin $j$, and we use an index-free notation where $\hat\Om_{2,j}(z,\omega) = \hat\Om_{2,j}(z)^{a_1 \ldots a_j} \omega_{a_1} \ldots \omega_{a_j}$ and $\omega_a$ is a null vector in the parallel directions. For brevity we define $\hat{\Delta}_{23} \equiv \hat{\Delta}_2 - \hat{\Delta}_3$, and $y^a$ is defined in~\eqref{eq:chiraldist}, where one should remember that $a = 0,1$ are the parallel coordinates. 
Conservation of $R$-symmetry implies $r_\phi + r_2 + r_3 = 0$.
The function $\mathcal{F}^{\text{3pt}} (\chi)$ depends on the superconformal three-point invariant $\chi$. Like in the two-point function case, there is a unique, non-nilpotent, three-point invariant:
\begin{equation}\label{eq:3ptinv02}
\chi =    \frac{|y^a|^2}{(y^\bot)^2}\,.
\end{equation}
The function $\mathcal{F}_{\text{3pt}} (\chi)$ can be expanded in three-point superconformal blocks which are in turn sums of three-point bosonic blocks (reviewed in appendix \ref{app:3pt_bos_blocks}). 
Notice that there is no crossing equation for this correlator. We can act with the boundary Casimir on point $z_1$
and obtain the eigenvalue equation
\begin{equation}\label{eq:caseq3pt02}
\hat{\mathcal{C}}_{\text{susy}}^{(1)} 
\langle \Phi(y, \theta) \hat{\Oop}_{2,j} (0,\omega) \hat{\Oop}_{3} (\infty) \rangle =  \hat{C}_{\hat{\Delta},j,\mathit{r}} \langle \Phi(y, \theta) \hat{\Oop}_{2,j} (0,\omega) \hat{\Oop}_{3} (\infty) \rangle  \,.
\end{equation}
By now the story is familiar; we give an ansatz in terms of bosonic blocks and obtain a solution with 
 $\mathit{r} = \mathit{r}_\phi$, $j_{\text{chiral}} = 0$, $\hat{\Delta} = \Delta_\phi = \mathit{r}_\phi$ :
\begin{equation}\label{eq:threeptblocksusy02}
\hat{F}^{\text{3pt}}_{\Delta_\phi} ( \chi) = \hat{f}^{\text{3pt},\hat{\Delta}_{23}}_{\Delta_\phi,j} (\chi)\,,
\end{equation}
which describes the exchange of a boundary chiral field.
The other possible solution has $r = r_\phi-1$, $j_{\text{fermi}} = \half$ and generic $\hat\Delta$ :
\begin{equation}\label{eq:threeptblocksusy02Ferm}
\hat{F}^{\text{3pt}}_{\hat\Delta} (\chi) = \hat{f}^{\text{3pt}, \hat{\Delta}_{23}}_{\hat\Delta+\half,j} (\chi) \, ,
\end{equation}
and corresponds to the exchange of a Fermi multiplet.
Let us also consider the case where the second operator is a Fermi field. The three-point function is given by
\begin{equation}
\label{eq:three-PhiPsiO3}
\langle\Phi(y,\theta) \hat{\bar{\Psi}} (0) \hat{\mathcal{O}} (\infty) \rangle = \frac{(\gamma_{1 \beta})^{\mu} y_{\mu} \theta^{\beta}}{(y^\bot)^{\Delta   +  \hat{\Delta}_{23} + \frac{3}{2}}} \mathcal{F}_{\text{3pt}} (\chi)\,,
\end{equation}
where $\chi$ is the same invariant as before. 
There are again two solutions to the eigenvalue equation, the first one corresponds to the exchange of a boundary chiral 
\begin{equation}
\hat{F}^{\text{3pt}}_{\Delta_\phi} ( \chi) =  \hat{f}^{\text{3pt},\hat{\Delta}_{23} + \half}_{\Delta_\phi,0} ( \chi) \,,
\end{equation}
while the second describes a Fermi field  
\begin{equation}\label{eq:3ptsusyblockfermi02}
\hat{F}^{\text{3pt}}_{\hat{\Delta}} ( \chi) =  f^{\text{3pt},\hat{\Delta}_{23} + \half}_{\hat{\Delta} + \half,0} (\chi) \,.
\end{equation}
The supersymmetric block corresponds to a bosonic block with shifted external conformal dimensions $\hat{\Delta}_{23} \to \hat{\Delta}_{23} + \half$ and with spin $j=0$. Like in the two-point case, the shift can be understood as a contribution coming from a superconformal descendant of $\hat{\bar\Psi}$.
%
%
\subsubsection{Free theory in the bulk}\label{sec:freebulk02}
Having obtained a handful of correlators, let us investigate the possible constraints that a free theory in the bulk imposes on the boundary data. In superspace the free field equations of motion take the form
\begin{equation}
D^\alpha D_\alpha \Phi(y, \theta) = 0\,,
\end{equation}
which is the supersymmetric version of the more familiar $ \partial^2 \phi(x)=0$.
As usual, a free chiral field has dimension $\Delta_\phi = r_\phi = \frac{1}{2}$.
Imposing this condition on the two bulk-to-boundary correlators \eqref{eq:bulkbdychir02} and \eqref{eq:bulkbdyfermi02}, we obtain two solutions:
\begin{equation}
\label{eq:free_bulk_bdy_20}
\langle \Phi (y,\theta) \hat{\bar{\Phi}}_{\hat\Delta=\half} (0) \rangle = \frac{b_{\Phi \hat{\bar{\Phi}}}}{ |y^\mu|}\, ,
\qquad
\langle\Phi(y, \theta) \hat{\bar{\Psi}}_{\hat{\Delta} = 1} (0) \rangle = \frac{b_{\Phi \hat{\bar{\Psi}}} \gamma_{1 \beta}^\mu y_{\mu} \theta^\beta }{|y^\mu|^{3}}\,.
\end{equation}
This is not surprising. The first solution corresponds to a boundary chiral field of dimension $\hat{\Delta} =  \half$, which corresponds to the operator  $\hat{\phi}$ and describes Neumann boundary conditions. The second solution
is a Fermi field with  $\hat{\Delta} = 1$, which has a scalar descendant with dimension $\hat{\Delta}+\half = \frac{3}{2}$ (recall the discussion below \eqref{eq:bulkbdyfermi02}). The descendant can be identified with $\partial_\bot \hat{\phi}$ as expected for Dirichlet boundary conditions. We have therefore proven that the boundary expansion of a bulk free field has a finite number of contributions.

We now turn to the three-point function to see if there are extra constraints on the boundary operators from a free bulk chiral field. 
Let us expand the correlation function~\eqref{eq:three-pt-PO2O3} in bosonic blocks, where we take $\Phi$ to be a free bulk chiral.
From equation~\eqref{eq:free_bulk_bdy_20} we know that there are two independent contributions coming from a chiral and a Fermi boundary field:
\begin{align}
\label{eq:freebulkbdychir02}
\Fm^{\text{3pt}} (\chi)
 = b_{\phi\hat\phi} \lambda_{\hat\phi\hat O_2 \hat O_3} 
   \hat{f}^{\text{3pt},\hat{\Delta}_{23}}_{\hat{\Delta} = \half,j} (\chi) 
 + b_{\phi \partial_\bot \hat\phi} \lambda_{\hat \partial_\bot \hat\phi \hat{O}_2 \hat O_3} 
   \hat{f}^{\text{3pt},\hat{\Delta}_{23}}_{\hat{\Delta} = \frac{3}{2},j}(\chi) \, .
\end{align}
Note that we have written the OPE coefficients explicitly in terms of the operators that appear in the OPE and not in terms of the superprimaries.
Equation~\eqref{eq:freebulkbdychir02} is identical to the conformal block expansion of a non-supersymmetric free scalar in the bulk, which has been studied in detail in~\cite{Lauria:2020emq, Behan:2020nsf}. 
In the limit $\chi \to 0$ there are unphysical singularities, which can only be removed provided the OPE coefficients satisfy the following relation:
\begin{align}
\label{eq:ope-rels}
 b_{\phi \partial_\bot \hat\phi} 
 \lambda_{\hat \partial_\bot \hat\phi \hat{O}_2 \hat O_3} 
 = -
 \frac{2 
    \Gamma \! \left(\frac{2 j - 2 \hat\Delta_{23} + 3}{4} \right) 
    \Gamma \! \left(\frac{2 j + 2 \hat\Delta_{23} + 3}{4} \right)}{
    \Gamma \! \left(\frac{2 j - 2 \hat\Delta_{23} + 1}{4} \right) 
    \Gamma \! \left(\frac{2 j + 2 \hat\Delta_{23} + 1}{4} \right)}
 b_{\phi\hat\phi} \lambda_{\hat\phi\hat O_2 \hat O_3} \, .
\end{align}
This constraint is equivalent to the constraints on non-supersymmetric three-point functions with a free bulk. We can go one step further and look at the three-point function involving a boundary Fermi multiplet~\eqref{eq:three-PhiPsiO3} in the hope that we will find additional constraints on the CFT data from supersymmetry.
Once again, we expect the two solutions in equation~\eqref{eq:free_bulk_bdy_20} to contribute to the Fermi three-point function. 
If we act with the equations of motion, the resulting differential equation can only be solved if $\hat \Delta_{23} = 1$, excluding the solution $\hat{\Delta} = \half$. The resulting correlator corresponds to a single bosonic block
\begin{align}
\label{eq:freebulkbdyfermi02}
\Fm_{\text{3pt}}(\chi) 
\propto \hat f^{\text{3pt},\frac{3}{2}}_{\hat\Delta=\frac{3}{2},0}(\chi) 
= \frac{1}{(\chi +1)^{\frac{3}{2}}} \, .
\end{align}
In this case the correlator is manifestly non-singular as $\chi \to 0$.
Having a free bulk implies that there is only one operator in the OPE $\hat \phi \times \hat \psi$, which has fixed dimension $\hat \Delta_3 = \hat \Delta_2 - 1$. This is a new, additional constraint coming from the superspace analysis that was not present in the non-supersymmetric case. It would be interesting to see if a more systematic analysis allows us to find more general constraints.
%
%
\subsection{The $\mathcal{N} = (1,1)$ boundary }
We now present the superspace analysis for the $\Nm=(1,1)$ boundary, and since it is quite similar to what we have done so far, we will mostly state the results.
We again divide the superspace into parallel and perpendicular coordinates, the bosonic coordinates are split as usual, and for the fermionic variables we define
\begin{align}
\text{parallel:} \quad  \tilde{\theta}^{1} \equiv \tilde{\theta}^{-} = - i (\theta^{-} - \bar{\theta}^{-} )\,, &\quad \tilde{\theta}^{2} \equiv \tilde{\theta}^{+} = - (\theta^{+} + \bar{\theta}^{+})\,, \\
\text{perpendicular:} \quad  \theta_{\bot}^{1} \equiv \theta_{\bot}^{-} = - (\theta^{-} + \bar{\theta}^{-})\,, &\quad \theta_{\bot}^{2} \equiv \theta_{\bot}^{+} = - i (\theta^{+} - \bar{\theta}^{+})\,,
\end{align}
There are two useful ways to construct supersymmetric perpendicular distances
\begin{equation}\label{eq:ppdist11}
     z^\bot =      y^\bot + 2 i \theta^- \theta^+\,, \qquad
\bar z^\bot = \bar y^\bot + 2 i \bar \theta^- \bar \theta^+\,,
\end{equation}
with the property that they are chiral and antichiral respectively $\bar D_\a z^\bot = D_\a \bar z^\bot = 0$.
These distances will be the natural objects to appear in correlators of  (anti)chiral  fields.
The decomposition of a bulk (anti)chiral field for the $\mathcal{N} = (1,1)$ boundary contains only one boundary supermultiplet instead of the two possibilities present in the $(0,2)$ boundary
\begin{equation}
\Phi = \hat{\Phi} + \ldots \, ,
\end{equation}
where the dots stand for derivatives of $\hat{\Phi}$. The field $\hat{\Phi}$ can be decomposed into bosonic components, schematically (see \cite{Brunner:2019qyf}  for the precise coefficients)
\begin{equation}
\hat{\Phi} = \hat{\phi} + \tilde{\theta}^{+} \hat{\psi}_{+} + \tilde{\theta}^{-} \hat{\psi}_{-} + \tilde{\theta}^2 \partial_\bot \hat{\phi} + \tilde{\theta}^{2} \hat{F} \, .
\end{equation}
We see that $\hat{\phi}$ and $\partial_\bot \hat{\phi}$  belong to the same boundary multiplet, which implies the unexpected feature that Neumann and Dirichlet boundary conditions are related by supersymmetry. 
%
\subsubsection{One-point functions}
Due to the absence of $R$-symmetry, (anti)chiral bulk fields can now acquire a one-point function. The only correlators consistent with the symmetry constraints are given by
\begin{equation}
\langle \Phi (y, \theta) \rangle 
= \frac{a_\Phi}{(2 z^{\bot})^{\Delta}}\,, \qquad 
\langle \bar{\Phi} (\bar{y},\bar\theta) \rangle 
= \frac{a_{\bar\Phi}}{(2 \bar{z}^{\bot})^{\Delta}}\,,
\end{equation}
where $z^\bot, \bar{z}^\bot$ were defined in~\eqref{eq:ppdist11}, and $a_\Phi$ is the one-point coupling that appears as the coefficient of the ``boundary identity'' in the conformal block expansion.
%
%
\subsubsection{Bulk-to-boundary correlator}
\label{sec:bulk_to_bdy11}
Since a chiral bulk supermultiplet decomposes into one boundary supermultiplet, we expect only one correlator:
\begin{equation}\label{eq:bulkbdychir11}
\langle\Phi(y, \theta) \hat{\Oop} (0) \rangle 
= \frac{b_{\Phi \hat{\Om}}}{(2 z^{\bot})^{\Delta - \hat{\Delta}} |y^\mu|^{2\hat{\Delta}}}\,,
\end{equation}
where $z^\bot$ is the same as above and $|y^\mu|$ is the norm of the chiral coordinate~\eqref{eq:chiraldist}. Notice that $\hat{\Delta}$ is unconstrained so these are the operators captured by the boundary conformal blocks to be calculated below.
%
%
\subsubsection{Two-point functions}
Due to the broken $R$-symmetry there is now no selection rule implying that correlators with fields of the same chirality vanish.
Thus, we should consider the two-point functions $\langle \Phi_1 \bar\Phi_2 \rangle$ and $\langle \Phi_1 \Phi_2 \rangle$ where the $R$-charges are arbitrary. 
The two-point functions in the presence of the $\mathcal{N} = (1,1)$ boundary have the same structure as in the $\mathcal{N} = (0,2)$ case.
Each of them depends on a single superconformal invariant $\xi$ which has the appropriate chirality properties:
\begin{align}
& \langle \Phi_1(y_1,\theta_1) \bar \Phi_2(\bar y_2, \bar\theta_2) \rangle 
= \frac{\mathcal{F}^{\phi \bar{\phi}} (\xi)}
       {(2 z_1^\bot)^{\Delta_1} (2\bar{z}_{2}^\bot)^{\Delta_2}}, \qquad
\xi = \frac{(y_{12})^2}{4 z_{1}^\bot \bar{z}_{2}^\bot}\,, \\[0.5em]
& \langle \Phi_1(y_1,\theta_1) \Phi_2(y_2,\theta_2) \rangle  
= \frac{\mathcal{F}^{\phi \phi} (\xi)}
       {(2 z_{1}^\bot)^{\Delta_1} (2 z_{2}^\bot)^{\Delta_2}}, \qquad
\xi = \frac{(\tilde{y}_{12})^2 + 2i \theta_{12}^2 (z_1^\bot + z_2^\bot)}
        {4 z_{1}^\bot z_{2}^\bot}\,.
\end{align}
The perpendicular distances $z^\bot,\bar{z}^\bot$ are given in equation~\eqref{eq:ppdist11}, the chiral-antichiral distance $y_{12}^\mu$ can be found in~\eqref{eq:chiraldistmu}, and we have defined the following chiral-chiral distance:
\begin{align}
\begin{split}
\tilde{y}_{12}^0 
& = y_1^0 - y_2^0 - 2i (\theta_1^+\theta_2^+ - \theta_1^-\theta_2^-)\,, \\
\tilde{y}_{12}^1 
& = y_1^1 - y_2^1 - 2i (\theta_1^+\theta_2^+ + \theta_1^-\theta_2^-)\,, \\
\tilde{y}_{12}^2 
& = z_{1}^\bot - z_{2}^\bot\,.
\end{split}
\end{align}
Let us now calculate the corresponding superblocks for the functions $\mathcal{F}^{\phi \bar{\phi}} (\xi)$ and $\mathcal{F}^{\phi \phi} (\xi)$.
%
%
\subsubsection{Superconformal blocks}

We now calculate the superconformal blocks using the same approach we used in the $\Nm=(0,2)$ case in section \ref{sec:susyblocks02}. We use the Casimir to obtain a differential equation that we then solve using a finite combination of bosonic blocks.
\paragraph{Bulk channel:} We first act wit the bulk Casimir%
\begin{equation}\label{eq:bulkcaseqsusy11}
\mathcal{C}_{\text{susy}}^{(12)} 
\langle \Phi_1(y_1, \theta_1) \bar \Phi_2(\bar y_2, \bar \theta_2) \rangle = C_{\Delta,\ell,r} 
\langle \Phi_1(y_1, \theta_1) \bar \Phi_2(\bar y_2, \bar \theta_2) \rangle\,,
\end{equation}
where $\mathcal{C}_{\text{susy}}^{(12)}$ and $C_{\Delta,\ell,r}$ were already given in \eqref{eq:bulkcassusy} and \eqref{eq:bulkcaseigensusy} respectively. 
The solution to this equation in terms of bosonic blocks is easy to find.
Only $\ell = 0$ and $r = r_1 - r_2$ contributes
\begin{align}\label{eq:bulkblock11}
\begin{split}
F^{\phi_1\bar\phi_2}_{\Delta} (\xi) 
& =  f^{\Delta_{12}}_{\Delta} (\xi) 
+ \frac{(\Delta-\Delta_{12})(\Delta+\Delta_{12})}{(2 \Delta -1) (2 \Delta +1)}  f^{\Delta_{12}}_{\Delta + 2} (\xi)\,\\
 & = \xi^{\frac{\Delta -\Delta_1-\Delta_2}{2}}  \,  _2F_1\Big(\frac{\Delta - \Delta_{12} }{2},\frac{\Delta + \Delta_{12} }{2}; \Delta +\frac{1}{2}; - \xi \Big),
\end{split}
\end{align}
which in general corresponds to a long operator being exchanged in the $\phi_1 \times \bar \phi_2$ OPE.
The contributions of short operators can be found by evaluating $\Delta$ at the unitarity bound, as discussed below equation~\eqref{eq:bulkblocks02extra}.
For the two-point function $\langle \Phi_1 \Phi_2 \rangle$, which was not present in the $\Nm=(0,2)$ case, the solution to the Casimir equation
\begin{equation}\label{eq:bulkcaseqsusyff11}
\mathcal{C}_{\text{susy}}^{(12)} 
\langle \Phi_1(y_1,\theta_1) \Phi_2(y_2,\theta_2) \rangle = C_{\Delta,\ell,r} 
\langle \Phi_1(y_1,\theta_1) \Phi_2(y_2,\theta_2) \rangle\,,
\end{equation}
can be written in terms of single bosonic blocks with shifted arguments $F^{\phi\phi}_{\Delta} = f_{\Delta+\frac{n}{2}}$.
This is a well-known result which has been described in detail for $d=3$ in~\cite{Bobev:2015jxa}. 
We will review the analysis in detail in section~\ref{sec:bulk-4eps}.
\paragraph{Boundary channel:} Let us now move on to the boundary channel. The boundary Casimir is now given by
\begin{equation}
\hat{\mathcal{C}}_{\text{susy}} 
=  - \mathcal{D}^{2} -  \half \left\{ \mathcal{K}^{a} ,\mathcal{P}_{a} \right\}  + \half \mathcal{M}^{a b} \mathcal{M}_{ab }  
+ \frac{1}{4}  [\tilde{\Sop}^\a, \tilde{\Qop}_\a]\,,
\end{equation}
with eigenvalue 
\begin{equation}
\label{eq:bdycaseigensusy11}
\hat{C}_{\hat{\Delta}, j} 
= \hat{\Delta} (\hat{\Delta} - 1) + j^2\,.
\end{equation}
We can only find consistent solutions when the superprimary has no parallel spin: $j=0$.
For the chiral-antichiral correlator we find
\begin{align}
\begin{split}\label{eq:phiphiBarbdyblocks11}
\hat{F}^{\phi_1 \bar{\phi}_2}_{\hat{\Delta}} (\xi) &= \hat{f}_{\hat{\Delta}} (\xi) + \frac{1}{4} \hat{f}_{\hat{\Delta} + 1} (\xi)\\
&= \xi^{-\hat{\Delta} } \, _2F_1\Big(\hat{\Delta} -\frac{1}{2}, \hat{\Delta} ;2 \hat{\Delta} ; -\frac{1}{\xi}\Big)\,.
\end{split}
\end{align}
while for the chiral-chiral correlator we have
\begin{align}\label{eq:phiphibdyblocks11}
\begin{split}
\hat{F}^{\phi_1 \phi_2}_{\hat{\Delta}} (\xi) &= \hat{f}_{\hat{\Delta}} (\xi) - \frac{1}{4} \hat{f}_{\hat{\Delta} + 1} (\xi) \\
&= \xi^{-\hat\Delta} \,_2F_1 \Big( \hat\Delta + \half, \hat\Delta; 2\hat\Delta; - \frac{1}{\xi}\Big)\,.
\end{split}
\end{align}
These two blocks describe the exchange of operators whose correlator \eqref{eq:bulkbdychir11} is non-vanishing.
This concludes our analysis of two-point blocks in the $\Nm=(1,1)$ boundary. We will generalize these results for arbitrary $3 \leq d \leq 4$ in section \ref{sec:4-eps}. The superspace analysis of this section will give supporting evidence that the blocks of section \ref{sec:4-eps} are a consistent continuation of the $3d$ results presented here.

%
%
\subsubsection{Three-point functions}
Let us now study the correlator of a chiral bulk field and two boundary fields. We allow the first boundary operator to have arbitrary spin $j$, and we will work in a frame where we set $x_{2}^a, \tilde{\theta}_{2}^a, \tilde{\theta}_{3}^a $ to zero, and $x_{3}^a$ to infinity. 
Unlike the situations studied so far, there is a nilpotent invariant consistent with all the symmetries, which implies the following structure
\begin{equation}\label{eq:threeptsusy11}
\langle \Phi(y,\theta) \hat{\Oop}_{2,j} (0, \omega) \hat{\Oop}_{3} (\infty) \rangle =  \frac{\left(y^a \omega_a\right)^j} {(y^{\bot})^{\Delta_\phi + \hat{\Delta}_{23}} |y^a|^j} \left(
    \mathcal{F}^{\text{3pt}}_{1}(\chi)
    + \frac{\theta^+\theta^-}{y^\bot}
      \mathcal{F}^{\text{3pt}}_{2}(\chi)
\right).
\end{equation}
All the dependence of the correlator is in terms of the chiral coordinates $y$ and $\theta$, see~\eqref{eq:chiraldist}.
The superconformal invariant $\chi$ is the same as for the $\Nm = (0,2)$ boundary in~\eqref{eq:3ptinv02}.
The superfields $\Phi, \hat{\Oop}_i$ appearing in the three-point function can be expanded into bosonic components, whose correlators are captured by $\mathcal{F}^{\text{3pt}}_{i}$. Let us look at this expansion with more details. Since we chose a frame where $\theta_2 = \bar{\theta}_2 = \theta_3 = \bar{\theta}_3 = 0$, only the superprimary in the $\theta$-expansion of $\hat{\Oop}_i$ will contribute, then
\begin{align}
\label{eq:phiexp11}
\Phi(y,\theta)
= \phi (y) + \theta^\alpha \psi_\alpha(y) + \theta^+ \theta^- F(y)\,, \quad
\hat{\Oop}_2 (0, \omega) = \hat{O}_2(0, \omega)\,, \quad
\hat{\Oop}_3 (\infty) = \hat{O}_3(\infty)\,.
\end{align}
Comparing the correlator~\eqref{eq:threeptsusy11} with the expansion~\eqref{eq:phiexp11} we read off
\begin{align}
 \langle \phi(x) \hat O_2(0,\omega) \hat O_3(\infty) \rangle &
 = \frac{(x^a w_a)^j}{(x^{\bot})^{\Delta_\phi + \hat{\Delta}_{23}} |x^\mu|^{j}} \mathcal{F}^{\text{3pt}}_{1}(\chi)\,,  \\
 \langle \psi_{\alpha} (x) \hat O_2(0,\omega) \hat O_3(\infty) \rangle & = 0\,, \\
 \langle F(x) \hat O_2(0,\omega) \hat O_3(\infty) \rangle & 
 = \frac{(x^a w_a)^j }{(x^{\bot})^{(\Delta_\phi + 1) + \hat{\Delta}_{23}} |x^{\mu}|^{j}} \mathcal{F}^{\text{3pt}}_{2}(\chi)\,,
\end{align}
so indeed $\Fm^{\text{3pt}}_{1,2}$ capture the correlators of the top and bottom components of the chiral multiplet.
To find the corresponding superconformal blocks we act with the boundary supersymmetric Casimir in point $z_1$:
\begin{equation}
\hat{\mathcal{C}}^{(1)}_{\text{susy}} 
\langle \Phi(y,\theta) \hat\Oop_{2,j} (0, \omega) \hat\Oop_3 (\infty) \rangle 
= \hat{C}_{\hat{\Delta},0,\mathit{r}} 
\langle \Phi(y,\theta) \hat\Oop_{2,j} (0, \omega) \hat\Oop_3 (\infty) \rangle\,,
\end{equation}
where $\hat{C}_{\hat{\Delta},0,\mathit{r}}$ is given in equation~\eqref{eq:bdycaseigensusy11}. This results in two coupled differential equations, which we can solve by assuming that the superconformal blocks are given in terms of the bosonic blocks $\hat{f}^{\text{3pt}}_{\hat{\Delta}}$ given in~\eqref{eq:threeptbosblocks}. The final result reads
\begin{align}\label{eq:threeptblocks11}
F^{\text{3pt}}_{1,\hat{\Delta}}(\chi) &= f^{\text{3pt},\hat{\Delta}_{23}}_{\hat{\Delta},j} (\chi) + c_{\hat\Delta} f^{\text{3pt},\hat{\Delta}_{23}}_{\hat{\Delta} + 1,j}  (\chi)\,, \nonumber \\
F^{\text{3pt}}_{2,\hat{\Delta}} (\chi) 
&= -2i(\mathit{r}_\phi- \hat{\Delta}) f^{\text{3pt},\hat{\Delta}_{23}}_{\hat{\Delta},j} (\chi) 
 +  2 i c_{\hat\Delta} (\mathit{r}_\phi + \hat{\Delta} - 1) f^{\text{3pt},\hat{\Delta}_{23}}_{\hat{\Delta} + 1,j} (\chi)\,,
\end{align}
where $c_{\hat\Delta}$ is a free parameter, related to the OPE coefficients of the exchanged operator, see equation~\eqref{eq:expansion-3pt-11} below.

\subsubsection{Free bulk theory}
We now repeat the analysis of section \ref{sec:freebulk02}, and see how the bulk equations of motion constrain the spectrum of boundary operators. 
Imposing that the chiral field is free in \eqref{eq:bulkbdychir11} fixes the dimension of the boundary field to 
$\hat{\Delta} = \half$. Unlike in the $\mathcal{N} = (0,2)$ case there is only solution, since both Neumann and Dirichlet boundary conditions are related by supersymmetry, and belong to the same supermultiplet. 

Let us now focus on to the three-point function \eqref{eq:threeptsusy11}. It is well known that the free equations of motion for a chiral field imply $F(x) = 0$, so it is sufficient to focus on $\Fm_{1}^{\text{3pt}} (\chi)$. 
From the analysis of the free bulk-to-boundary correlator we conclude that there can only be one multiplet in the bulk-to-boundary OPE. The superprimary has dimension $\hat\Delta = 
\half$, which we will call $\hat\phi$. The multiplet also contains a superdescendant of dimension $\hat\Delta + 1 = \frac{3}{2}$, which we denote by $\partial_\bot\hat\phi$.
Both operators contribute to the superconformal block~\eqref{eq:threeptblocks11}, and the resulting correlation function is
\begin{align}\label{eq:expansion-3pt-11}
 \Fm_{1}^{\text{3pt}} (\chi)
 = b_{\phi\hat\phi} \lambda_{\hat\phi\hat O_2 \hat O_3} 
   \hat{f}^{\text{3pt},\hat{\Delta}_{23}}_{\hat{\Delta} = \half,j} (\chi) 
 + b_{\phi \partial_\bot \hat\phi} \lambda_{\hat \partial_\bot \hat\phi \hat{O}_2 \hat O_3} 
   \hat{f}^{\text{3pt},\hat{\Delta}_{23}}_{\hat{\Delta} = \frac{3}{2},j}(\chi) \, ,
\end{align}
where the OPE coefficients are written in terms of the superdescendants, not the superprimaries.
Due to the presence of a free coefficient $c_{\hat{\Delta}}$ in the superconformal block~\eqref{eq:threeptblocks11}, the relative coefficient in this expansion is not fixed by supersymmetry.
Equation~\eqref{eq:expansion-3pt-11} is identical to the non-supersymmetric case of a free scalar in the bulk and to equation~\eqref{eq:freebulkbdychir02} for the $\Nm = (0,2)$ boundary.
Thus, the analysis below~\eqref{eq:freebulkbdychir02} applies here as well and we find the same OPE relations~\eqref{eq:ope-rels}. There are no extra constraints coming from supersymmetry.

\section{Boundaries across dimensions}
\label{sec:4-eps}

In this section we study superconformal theories with boundaries in any, in principle continuous, number of dimensions $3 \le d \le 4$, keeping the codimension fixed.
We obtain superconformal blocks using similar techniques as were developed originally for bulk four-point functions in~\cite{Bobev:2015jxa,Bobev:2017jhk}.\footnote{Another example of blocks across dimensions was uncovered in the context of Parisi-Sourlas supersymmetry \cite{Kaviraj:2019tbg,Kaviraj:2020pwv}.}
Conformal blocks in an arbitrary number of dimensions allow us to use analytical techniques like the $\epsilon$-expansion, a subject that we explore in this section inspired by previous work \cite{Liendo:2012hy,Bissi:2018mcq} .

\subsection{Superconformal blocks}

\subsubsection{Superconformal algebra}

In the entire section we follow the same conventions as~\cite{Bobev:2015jxa}, which we review briefly. The notation will differ from the one in section~\ref{sec:3d}, but our main results, the superconformal blocks, will be convention-independent. We hope this does not cause too much confusion. The reader is welcome to look at the original reference for more details.
The conformal part of the algebra is generated by the usual operators $D$, $P_i$, $K_i$ and $M_{ij}$. 
We also have four Poincaré supercharges $Q^+_\a$ and $Q^-_\ad$ and four conformal supercharges $S^{\ad+}$ and $S^{\a-}$ with anticommutation relations 
\begin{align}
 \{ Q^+_\a, Q^-_\ad \} = \Sigma^i_{\a\ad} P_i\,, \qquad
 \{ S^{\ad+}, S^{\a-} \} = \bar \Sigma_i^{\ad\a} P_i\,, \qquad
 i = 1, \ldots, d\,.
\end{align}
Finally, there is a generator $R$ of $U(1)_R$ symmetry, under which $Q^+_\a$ and $Q^-_\ad$ have charge $+1$ and $-1$ respectively.
Provided that $\Sigma^i_{\a\ad}$ satisfies certain formal identities, the superjacobi identites are satisfied for arbitrary $d$.
The full set of commutation relations, the Casimir operator $C_{\text{bulk}}$, and many other important relations can be found in~\cite{Bobev:2015jxa}.

In what follows, we will focus our attention on chiral primary operators $\phi$ and their complex conjugates $\bar \phi$.
These operators are killed by supercharges of the same chirality, and using the superconformal algebra their conformal dimension is related to the $R$-charge:
\begin{align}
   \left[ Q^+_\a, \phi(0) \right] 
 = \left[ Q^-_\ad, \bar\phi(0) \right]
 = 0 \quad \Rightarrow \quad
   \Delta_{\phi} 
 = \Delta_{\bar\phi} 
 = \frac{d-1}{2} r_\phi
 = - \frac{d-1}{2} r_{\bar\phi}\,.
\end{align}
The chirality property, as well as the relation between $\Delta$ and $r$, will be important in the calculation of superconformal blocks in the next section.

The subalgebra of conformal transformations that preserve the boundary is generated by $D$, $P_a$, $K_a$ and $M_{ab}$, where $a,b = 2, \ldots, d$.
We chose $P_1$ not to be part of this subalgebra, which physically means that the boundary sits at $x_1 \equiv x^\bot = 0$. 
Only half of the original supercharges belong to the algebra, and they anticommute as:
\begin{align}
 \{ Q^{\text{bdy}}_A, Q^{\text{bdy}}_B \} = (\Sigma_{bdy})_{AB}^a P_a\,, \qquad
 \{ S^{\text{bdy}}_A, S^{\text{bdy}}_B \} = (\Sigma_{bdy})_{AB}^a K_a\,, \qquad 
 A, B = 1,2.
\end{align}
For arbitrary $d$ we embbed the boundary subalgebra into the full superconformal algebra as
\begin{align}
\label{eq:bdy-gens}
 Q^{\text{bdy}}_1 = Q^+_1 + Q^-_2\,, \quad
 Q^{\text{bdy}}_2 = Q^+_2 + Q^-_1\,, \quad
 S^{\text{bdy}}_1 = S^+_2 + S^-_1\,, \quad
 S^{\text{bdy}}_2 = S^+_1 + S^-_2\,.
\end{align}
It is easy to check explicitly in $d = 3$ and $d=4$ that~\eqref{eq:bdy-gens} indeed generate a subalgebra and that all the superjacobi identities are satisfied, provided that we use the following Clifford algebra representation:
\begin{align}
\label{eq:Clifford}
 \Sigma^i_{\a\ad} 
 = (\bar \Sigma_i^{\ad\a})^* = (\sigma_3, \sigma_1, \sigma_2, i \mathds{1})\,.
\end{align}
Notice that the generator $R$ is not part of the boundary superalgebra.
In physical terms the $R$ charge is not conserved near the boundary, and both $\langle \phi_1 \phi_2 \rangle$ and $\langle \phi_1 \bar \phi_2 \rangle$ are non-vanishing two-point functions for any $r_{1,2}$.
These two-point functions have different superconformal block decompositions that we treat separately in the next section.

In order to compute superconformal blocks, we will need the explicit form of the superconformal Casimir of the boundary superalgebra:
\begin{align}
\label{eq:def-cas}
 C_{\text{bdy}} = 
  - D^2
  - \frac12 \{ P_{a}, K^{a} \}
  + \frac12 M_{ab} M^{ab}
  + \frac14 [ S^{\text{bdy}}_A, Q^{\text{bdy}}_A]\,.
\end{align}
If we consider a boundary operator with quantum numbers $\hat{\Delta}, j$, then it will be an eigenstate of the superconformal Casimir with eigenvalue
\begin{align}
\label{eq:def-eigval}
 \hat C_{\hat\Delta,j}
 = \hat\Delta (\hat\Delta - d + 2)
 + j(j- d + 3) \,.
\end{align}

\subsubsection{Boundary channel}

As discussed at length in the superspace section, the boundary channel blocks for a two-point function are eigenfunctions of the boundary superconformal Casimir~\eqref{eq:def-cas}.
We can naturally split the Casimir operator into a non-supersymmetric piece and a contribution coming from supersymmetry:
\begin{align}
 C_{\text{bdy}} = C_{\text{bdy,non-susy}} + C_{\text{bdy,susy}}\,, \qquad
 C_{\text{bdy,susy}} \equiv \frac{1}{4}[S^{\text{bdy}}_A, Q^{\text{bdy}}_A]\,.
\end{align}
We worked out the non-supersymmetric contribution in equation~\eqref{eq:bdycasbos}. 
Focusing only on the supersymmetric part and using the anticommutation relations we obtain:
\begin{align}
\begin{split}
 [C_{\text{bdy,susy}}, \phi_1(x)] |0 \rangle
 & = \left(
    \frac{d-1}{2} \left[ R, \phi_1(x) \right]
    - \frac{1}{2} \left\{ Q_2^-, \left[S^{1-}, \phi_1(x)\right] \right\} 
 \right) | 0 \rangle \\
 & = \Big(
    \Delta_1 \phi_1(x)
    + i x^\bot \{ Q_1^-, [Q_2^-, \phi_1(x)] \} 
 \Big) | 0 \rangle\,.
\end{split}
\end{align}
In appendix~\ref{sec:app-blocks-4-eps} we use superconformal Ward identities to rewrite the piece with $Q_1^- Q_2^-$ as a term that can be included in a differential equation. 
Unfortunately, we have not been able to find a strategy to use these Ward identities for general $d$.
Instead, we focus on the particular cases of $d=3,4$ where the explicit Clifford algebra representation~\eqref{eq:Clifford} is valid.
Since the final result does not depend on $d$, we claim it is also valid for $3 \le d \le 4$.\footnote{It is likely that our blocks are valid for $2 \le d \le 4$ but we have not checked explicitly the $d = 2$ case. Notice that below $d \le 3$ on has to take into account the operators $M_{\hat i, \hat j}$ with $\hat i, \hat j = d, \ldots, 4$, and the calculation is slightly more complicated.}
The fact that we can find solutions to the Casimir equations with the expected properties for any continuous $d$ confirms that our assumption is justified. The $\epsilon$-expansion results, to be described below and in the next section, also give supporting evidence that the whole picture is consistent.

\paragraph{\texorpdfstring{$\langle \phi_1 \bar \phi_2 \rangle$}{< phi1 phiB2>} correlator:}

When we consider the two-point function of a chiral and antichiral operator, the contribution from supersymmetry is given by
\begin{align}
\label{eq:bdy-contrib-PPb}
 \frac{C_{\text{bdy,susy}} \langle \phi_1(x_1) \bar \phi_2(x_2) \rangle}
      {(2x_1^\bot)^{-\Delta_1} (2 x_2^\bot)^{-\Delta_2}}
 = -\xi \partial_\xi \hat F_{\hat \Delta}^{\phi_1\bar\phi_2}(\xi)\,.
\end{align}
Combining the supersymmetric and non-supersymmetric pieces, and using the appropriate value of the Casimir, we get the following differential equation:
\begin{align}
 \left[
 \xi  (\xi +1) \partial_\xi^2
 + \left(\frac{d}{2} + (d -1) \xi \right) \partial_\xi
 - \big( \hat\Delta (\hat\Delta - d + 2) + j(j- d + 3) \big) \right]
 \hat F_{\hat \Delta}^{\phi_1\bar\phi_2}(\xi)
 = 0\,.
\end{align}
A priori, there are two independent solutions of this equation for arbitrary values $\hat \Delta$ and $j$.
However, we must also require that the solutions can be decomposed into non-supersymmetric blocks, and we find that this is only possible whenever $j = 0$ for arbitrary $\hat \Delta$.
The solution can be expressed either as a linear combination of bosonic blocks, or as a single hypergeometric function with a prefactor:
\begin{align}
\label{eq:bdy-block-PPb}
\begin{split}
 \hat F_{\hat\Delta}^{\phi_1\bar\phi_2}(\xi) 
 & =\hat f_{\hat\Delta}(\xi) 
 + \frac{\hat\Delta }{2(2\hat\Delta - d + 3)} \hat f_{\hat\Delta+1}(\xi)\,,\\
 & = \xi ^{-\hat\Delta } \, _2F_1\Big(
   \hat\Delta ,\hat\Delta + 1-\frac{d}{2} ;2 \hat\Delta-d+3;-\frac{1}{\xi }\Big) \,.
\end{split}
\end{align}
Even though we considered a general two-point function $\langle \phi_1 \bar \phi_2 \rangle$, the superconformal blocks are the same as for a two-point function of identical (anti)chiral operators $\langle \phi \bar \phi \rangle$.
A nice consistency check is that the relative coefficient between the non-supersymmetric blocks is positive, as we expect in the defect channel of $\langle \phi \bar \phi \rangle$, because the coefficients that appear in the OPE are $|b_{\phi\hat\Om}|^2$.
When we restrict to $d = 3$ we find perfect agreement with the explicit superspace calculation~\eqref{eq:phiphiBarbdyblocks11}.

\paragraph{\texorpdfstring{$\langle \phi_1 \phi_2 \rangle$}{< phi1 phi2>} correlator:}

In a similar way, we can work out the Ward identities for the $\langle \phi_1 \phi_2 \rangle$ two-point function.
The new contribution to the Casimir equation is:
\begin{align}
\label{eq:bdy-contrib-PP}
 \frac{C_{\text{bdy,susy}} \langle \phi_1(x_1) \phi_2(x_2) \rangle}
      {(2x_1^\bot)^{-\Delta_1} (2 x_2^\bot)^{-\Delta_2}}
 = -(\xi + 1) \partial_\xi \hat F_{\hat \Delta}^{\phi_1\phi_2}(\xi)\,.
\end{align}
Combining the non-supersymmetric and supersymmetric pieces with the eigenvalue~\eqref{eq:def-eigval}, the Casimir equation reads
\begin{align}
 \left[
 \xi(\xi+1) \partial_\xi^2
 + \left( \frac{d-2}{2} + (d-1) \xi \right) \partial_\xi
 - \big( \hat\Delta (\hat\Delta - d + 2) + j(j- d + 3) \big) \right]
 \hat F_{\hat\Delta}^{\phi_1\phi_2}(\xi )
 = 0\,.
\end{align}
Once again, we only find physically acceptable solutions whenever $j = 0$:
\begin{align}
\label{eq:bdy-block-PP}
\begin{split}
 \hat F^{\phi_1\phi_2}_{\hat\Delta}(\xi)
 & = \hat f_{\hat\Delta}(\xi)
 - \frac{\hat \Delta}{2(2\hat\Delta-d+3)} \hat f_{\hat \Delta+1}(\xi)\,,
  \\
 & = \xi^{-\Delta} \, {}_2F_1 \Big(
 \hat \Delta, \hat\Delta + 2 - \frac{d}{2}; 2\hat\Delta -d+3; -\frac{1}{\xi}
 \Big)\,.
\end{split}
\end{align}
The decompositions into non-supersymmetric blocks in~\eqref{eq:bdy-block-PP} and~\eqref{eq:bdy-block-PPb} are identical up to a relative minus sign.
We know this must be the case, since the boundary OPE of $\langle \phi\phi\rangle$ contains $b_{\phi\hat\Om}^2$, which is not necesarily positive definite, but instead
$b_{\phi\hat\Om}^2 = \pm |b_{\phi\hat\Om}|^2$.
When we restrict to $d = 3$ we find perfect agreement with the explicit superspace calculation~\eqref{eq:phiphibdyblocks11}.

\subsubsection{Bulk channel}
\label{sec:bulk-4eps}

Now we proceed to calculate the blocks that appear in the bulk decomposition using the bulk Casimir.

\paragraph{\texorpdfstring{$\langle \phi_1 \bar \phi_2 \rangle$}{< phi1 phiB2 >} correlator:}

To obtain bulk channel blocks we act with the full Casimir once more focusing on the part that is new from supersymmetry:
\begin{align}
\begin{split}
 C_{\text{bulk,susy}} 
 & =
 - \frac{d-1}{2} R^2
 + \frac{1}{2} [S^{\ad+}, Q^-_\ad]
 + \frac{1}{2} [S^{\a-}, Q^+_\a]\,.
\end{split}
\end{align}
We can simplify the action of the superconformal Casimir using the commutation relations, the chirality properties of $\phi_1$ and $\bar \phi_2$, and equation (51) from~\cite{Bobev:2015jxa}:
\begin{align}
\begin{split}
 \left[ C_{\text{bulk,susy}}, \phi_1(x_1) \bar \phi_2(x_2) \right] |0\rangle
 & = i x_{12}^\mu \bar \Sigma_\mu^{\ad\a}
     \left[ Q^-_\ad,      \phi_1(x_1) \right] 
     \left[ Q^+_\a , \bar \phi_2(x_2) \right] |0\rangle \\
 & \quad + \left( 
     2 (\Delta_1 + \Delta_{2})
     - \frac{d-1}{4} r_{12}^2
  \right) \phi_1(x_1) \bar \phi_2(x_2) |0\rangle \,.
\end{split}
\end{align}
Here we assume $\phi_i$ has charge $r_i$, we define $r_{12} = r_1 - r_2$ and we use chirality to relate $\Delta_i = \half (d-1) r_i $.
We can use Ward identities to rewrite the $Q$-dependent part in a way that can be put in a Casimir equation.
After some algebra we get\footnote{We find it more convenient to work in terms of $G(\xi) = \xi^{(\Delta_1+\Delta_2)/2} F(\xi)$, but one can easily map the results between the two conventions.}
\begin{align}
\label{eq:blk-contrib-PPb}
 \frac{C_{\text{bulk,susy}} \langle \phi_1(x_1) \bar \phi_2(x_2) \rangle}
      {(2x_1^\bot)^{-\Delta_1} (2 x_2^\bot)^{-\Delta_2} \xi^{-(\Delta_1+\Delta_2)/2}}
 = \left( 4 \xi \partial_\xi 
 - \frac{d-1}{4} r_{12}^2 \right) G_{\Delta}^{\phi_1\bar\phi_2}(\xi)\,.
\end{align}
Now we can combine all the pieces to form the differential equation
\begin{align}
\begin{split}
  \bigg[ 4 \xi ^2 (\xi +1) \partial_\xi^2
& + 2\xi ( 2 \xi  - d + 4) \partial_\xi 
  - \Delta(\Delta -d+2) \\
& - \ell(\ell+d-2)
  - \frac{d-1}{4} (r_{12}^2 - r^2)
  - \Delta_{12}^2 \xi 
  \bigg] G_{\Delta}^{\phi_1\bar\phi_2}(\xi)
  = 0\,.
\end{split}
\end{align}
The superselection rules in the $\phi_1 \times \bar \phi_2$ OPE were worked out in four dimensions~\cite{Poland:2010wg} and in any $d$~\cite{Bobev:2015jxa}.
For our setup, they imply that only superprimaries with $r = r_{12}$ and $\ell = 0$ can appear\footnote{Superprimaries with $\ell \ge 1$ also appear in the OPE but they have zero one-point function, so they are not relevant in our analysis.}.
Indeed, we can solve the Casimir equation in this case to find:
\begin{align}
\label{eq:bulk-block-across-bdy}
\begin{split}
 G_\Delta^{\phi_1\bar\phi_2}(\xi) 
& = g^{\Delta_{12}}_\Delta(\xi) 
   + \frac{(\Delta-\Delta_{12})(\Delta+\Delta_{12})}
          {(2 \Delta -d +2) (2 \Delta -d +4)} g^{\Delta_{12}}_{\Delta+2}(\xi) \\
 & = \xi ^{\Delta /2} \, 
   _2F_1\Big(
      \frac{\Delta + \Delta_{12}}{2},
      \frac{\Delta - \Delta_{12}}{2};
      \Delta +2 -\frac{d}{2};-\xi \Big) \, .
\end{split}
\end{align}
For generic values of $\Delta$ these blocks capture the exchange of a long operator, while they can be interpreted as short operators when $\Delta$ saturates the unitarity bounds.
The classification of possible short multiplets in $d = 3,4$ is well known and can be found for example in~\cite{Dolan:2002zh,Cordova:2016emh}.

\paragraph{\texorpdfstring{$\langle \phi_1 \phi_2 \rangle$}{< phi1 phi2 >} correlator:}

It is well know that when the two operators are chiral the bulk blocks are equal to non-supersymmetric blocks.
The precise selection rules for $\phi_1 \times \phi_2$ are known~\cite{Bobev:2015jxa}, but we review them here for convenience:
\begin{itemize}
 \item Consider a superprimary $\Om$ that has $R$-charge $r = r_1 + r_2 - 2$ and dimension $\Delta$. The descendant $(Q^+)^2 \Om$ has charge $r_1+r_2$, dimension $\Delta+1$ and is killed by $Q^+_\a$, so it appears in the $\phi_1 \times \phi_2$ OPE.
 \item Alternatively, consider the chiral superprimary operator $(\phi_1\phi_2)$, with $r = r_1 + r_2$ and $\Delta = \Delta_1 + \Delta_2$. In this case the superprimary itself is exchanged in the OPE.
 \item Finally, consider an anti-chiral superprimary operator $\bar \Psi$ whose dimension is related to its charge and given by $\Delta = -\frac{d-1}{2} r = d-1-(\Delta_1 + \Delta_2)$.
 The descendant operator $(Q^+)^2 \bar\Psi$ is exchanged in the OPE.
\end{itemize}
In what follows, whenever we consider bulk channel $\phi\phi$ superconformal blocks, $\Delta$ will be the dimension of the actual exchanged operator, and not the dimension of the superprimary.

\subsection{An aside: codimension-two defects}
\label{sec:codimension-two}

In the present paper we are mostly concerned with boundaries that interpolate between $3 \le d \le 4$ models.
In the same way there exist codimension-two defects that interpolate between a line in $d=3$ and a surface in $d=4$.
A familiar example is the $3d$ Ising twist defect, which was studied using Feynman diagrams in $4-\epsilon$ dimensions~\cite{Gaiotto:2013nva} (see also \cite{Billo:2013jda} for a Monte-Carlo analysis in exactly $d=3$).
These results were later reproduced and generalized using analytic bootstrap technology~\cite{Liendo:2019jpu}.
Similar techniques should be applicable to half-BPS codimension-two defects in supersymmetric theories like the Wess-Zumino model. 
We plan to come back to this problem in the future, but for now we describe how the superconformal blocks can be obtained within our framework.

The notation in this subsection will be different from the rest of the section; we hope this does not cause confusion.
We insert the codimension-two defect at $x_i = 0$ for $i = 1,2$ and label the parallel directions as $x_a$ for $a = 3, \ldots, d$.
The defect will naturally preserve parallel translations and special conformal transformations $P_a, K_a$, dilatations $D$, as well as parallel and perpendicular rotations $M_{ab}, M_{ij}$.
The two-point function of local operators depends on two cross-ratios.
To study the defect channel it is convenient to use coordinates $(\chi, \phi)$, while the bulk channel simplifies using coordinates $(x,\bar x)$:\footnote{Our cross-ratios are related to the ones in~\cite{Lemos:2017vnx} by $z = 1-x$ and $\bar z = (1-\bar x)^{-1}$.}
\begin{align}
\begin{split}
 & \frac{|x_{12}^a|^2 + |x_1^i|^2 + |x_2^i|^2}{|x_1^i||x_2^i|}
 = \chi 
 = \frac{2 - x -\bar x}{\sqrt{(1-x)(1-\bar x)}} \,, \\
 & \frac{x_1^i x_2^i}{|x_1^i||x_2^i|}
 = \cos \phi 
 = \frac{2 -x - \bar x + x \bar x}{2\sqrt{(1-x)(1-\bar x)}} \,.
\end{split}
\end{align}
The non-supersymmetric as well as the supercoformal blocks are given below.

Besides the bosonic generators described above, a half-BPS defect preserves two Poincaré supercharges:
\begin{align}
 Q^{\text{defect}}_1 = Q^+_1 \,, \qquad
 Q^{\text{defect}}_2 = Q^-_1 \,, \qquad
 S^{\text{defect}}_1 = S^{1+} \,, \qquad
 S^{\text{defect}}_2 = S^{1-} \,.
\end{align}
Our system does not preserve $R$-symmetry or transverse rotations independently, but only a linear combination of them that we call twisted transverse rotations:\footnote{For the particular case of a line defect in $d=3$, the subalgebra has been written explicitly in~\cite{Agmon:2020pde}.}
\begin{align}
\label{eq:twist-trans-spin}
 M^{\text{defect}} = M_{12} + \frac{d-1}{2} R \,.
\end{align}
With these conventions in mind, we proceed to obtain the superconformal blocks.

\subsubsection{Defect channel}

When supersymmetry is not present, the defect operators are labeled by the conformal dimension $\hat\Delta$ and the transverse spin $s$.
One can write down a Casimir equation which is solved by the following conformal blocks~\cite{Billo:2016cpy}:
\begin{align}
\label{eq:cod2-def-block}
 \hat f_{\hat\Delta,s}(\chi,\phi) 
 = e^{is\phi} \chi ^{-\hat\Delta } 
 \, _2F_1\Big(\frac{\hat\Delta }{2},\frac{\hat\Delta +1}{2};\hat\Delta+2-\frac{d}{2};\frac{4}{\chi ^2}\Big)\,.
\end{align}
In the supersymmetric case the only difference is that $s$ denotes the twisted transverse spin~\eqref{eq:twist-trans-spin}.
One can work out the selection rules, and find that only one operator in each multiplet contributes to the OPE, so the superconformal blocks are just~\eqref{eq:cod2-def-block} with the arguments shifted appropriately.

\subsubsection{Bulk channel}

Similarly, one can obtain a Casimir equation for the non-supersymmetric bulk channel.
It was observed in~\cite{Billo:2016cpy} that for codimension-two the Casimir equation is identical to the one found by Dolan and Osborn (D\&O) for bulk four-point functions~\cite{Dolan:2003hv}.
Therefore, the bulk-channel blocks of a defect two-point function are equal to the familiar four-point blocks:
\begin{align}
 f_{\Delta,\ell}(x, \bar x)
 = G^{0,0}_{\Delta,\ell,\text{D\&O}}(x, \bar x)\,.
\end{align}
When supersymmetry is included, the Casimir equation has an extra term $C_{\text{bulk,susy}}$ that can be simplified using Ward identites, as described in appendix~\ref{sec:app-blocks-4-eps}.
When the dust settles, it turns out that the blocks are described by non-supersymmetric blocks with shifted arguments:
\begin{align}
 F_{\Delta,\ell}(x, \bar x)
 = (x \bar x)^{-\half} G^{-1,-1}_{\Delta+1,\ell,\text{D\&O}}(x, \bar x)\,.
\end{align}
Even more surprisingly, these blocks are exactly the same that were found in~\cite{Bobev:2015jxa} for a four-point function of chiral and antichiral operators!

\subsection{Free theory}

After the small codimension-two detour let us come back to the boundary setup. As a first consistency check of our superconformal blocks, we consider a free chiral multiplet in the bulk in the presence of a half-BPS boundary.
It is well known that a free scalar has dimension $\Delta_\phi = \frac{d-2}{2}$, and the bulk equations of motion have a simple solution:
\begin{align}
 \partial_x^2 \langle \phi(x) \bar \phi(x') \rangle = 0
 \quad \Rightarrow \quad
 \Fm^{\phi\bar\phi}(\xi)
 = 
   \frac{c_1^{\phi\bar\phi}}{\xi^{(d-2)/2}}
 + \frac{c_2^{\phi\bar\phi}}{(\xi+1)^{(d-2)/2}}\,.
\end{align}
For the two-point function $\langle \phi \phi \rangle$  we find the same solution with free coefficients $c_{1,2}^{\phi\phi}$.
In order to impose supersymmetry, we must require that these correlators have consistent superconformal block decompositions in the bulk and boundary channels.
It is a simple exercise to show that this fixes $c_{1}^{\phi\phi} = c_{2}^{\phi\bar\phi} = 0$.
We can also fix $c_{1}^{\phi\bar\phi} = 1$ requiring that far away from the boundary, the two-point function $\langle \phi \bar \phi \rangle$ is unit normalized:
it is normalized such that the OPE coefficient of the bulk identity block is $1$.
Finally, after an appropriate redefinition $\phi \to e^{i\delta} \phi$ we can always chose the normalization $c_{2}^{\phi\phi} = 1$.
All in all, 
\begin{align}
\label{eq:free-sol-cross}
\begin{split}
 & \Fm^{\phi\bar\phi}(\xi)
 = \frac{1}{\xi^{(d-2)/2}}
 = F_{\text{Id}}^{\phi\bar\phi}(\xi)
 = \hat F_{(d-2)/2}^{\phi\bar\phi}(\xi)\,, \\
 & \Fm^{\phi\phi}(\xi)
 = \frac{1}{(\xi+1)^{(d-2)/2}}
 = F_{d-2}^{\phi\phi}(\xi)
 = \hat F_{(d-2)/2}^{\phi\phi}(\xi)\,.
\end{split}
\end{align}
In the above equation we also present the expansion of the correlation functions in terms of superconformal blocks.
Interestingly, only one superconformal block contributes to each channel, and with our normalization conventions all OPE coefficients are equal to one.

The above solution to crossing has a clear physical interpretation if we split the chiral primary operator in terms of its real and imaginary parts $\phi = \phi_1 + i \phi_2$.
Then we see that $\phi_1$ satisfies Neumann boundary conditions, whereas $\phi_2$ satisfies Dirichlet boundary conditions.
Indeed, from~\eqref{eq:free-sol-cross} we obtain
\begin{align}
 \lim_{x \to \text{bdy}} 
 \langle \partial_\bot \phi_1(x) \phi_1(x') \rangle = 0\,, \qquad
 \lim_{x \to \text{bdy}} \langle \phi_2(x) \phi_2(x') \rangle = 0\,.
\end{align}
We can think of our free correlation functions as linear combinations of the Neumann and Dirichlet boundary CFTs studied in~\cite{Liendo:2012hy}, with the precise relative coefficients fixed by supersymmetry.

\subsection{The $\epsilon$-expansion bootstrap}
\label{sec:eps-bootstrap}

It was originally observed in~\cite{Liendo:2012hy} that the crossing equation for boundary CFTs can be used to extract information about the Wilson-Fischer fixed point in the epsilon expansion.
In particular, they bootstrapped the one-loop correlators at order $\Om(\epsilon)$, and the analysis was generalized to $\Om(\epsilon^2)$ using different techniques in later works~\cite{Bissi:2018mcq, Kaviraj:2018tfd, Giombi:2020rmc}.
In this section we apply the same ideas to our supersymmetric two-point functions, and we obtain the full correlation functions at order $\Om(\epsilon)$.

In the supersymmetric setup there are two relevant crossing equations, one for $\phi\bar\phi$ and the other for $\phi\phi$:
\begin{align}
\label{eq:crossingEps}
 F^{\phi\bar\phi}_{\text{Id}}(\xi)
 + \sum_{n} c_n F^{\phi\bar\phi}_{\tilde\Delta_n}(\xi)
 = \sum_{n} \mu_n \hat F^{\phi\bar\phi}_{\hat \Delta_n}(\xi) \,, \quad
 \sum_{n} d_n F^{\phi\phi}_{\Delta_n}(\xi)
 = \sum_{n} \rho_n \hat F^{\phi\phi}_{\hat \Delta_n}(\xi) \,.
\end{align}
Notice that the spectrum of operators in the boundary channel is the same for the two correlators.
The boundary OPE coefficients are given in terms of bulk-to-boundary coefficients as $\mu_n = |b_{\phi\Om_n}|^2$ and $\rho_n = b_{\phi\Om_n}^2$, so they must be equal up to possible signs $\mu_n = \pm \rho_n$. 
The precise signs as a function of $n$ will be an outcome of our bootstrap analysis.
The bulk channel OPE coefficients are products of one- and three-point coefficients $c_n = a_{\Om_n} \lambda_{\phi\bar\phi\Om_n}$ and $d_n = a_{\Om_n} \lambda_{\phi\phi\Om_n}$ so we do not expect any relations between them.

Our analysis starts in $d = 4$, where the SCFT is free and the correlators are given in~\eqref{eq:free-sol-cross}.
We assume that the coupling of the theory is of order $g \sim \epsilon$, so as we lower the dimension to $d = 4-\epsilon$ the CFT data acquires small corrections.
In particular, we expect the external chiral operator to acquire an anomalous dimension:
\begin{align}
\begin{split}
  \Delta_\phi 
& = \frac{d-2}{2} 
  + \Delta_\phi^{(1)} \epsilon
  + \Delta_\phi^{(2)} \epsilon^2 
  + \ldots \, .
\end{split}
\end{align}
We should think of $\Delta_\phi^{(1)}$ as being related to the strenght of the coupling $g \propto \Delta_\phi^{(1)}\epsilon$, and the precise constant depends on the model under consideration.
In the bulk four-point function $\epsilon$-expansion bootstrap, see for example~\cite{Alday:2017zzv}, conservation of the stress tensor allows one to fix the precise value of $\Delta_\phi^{(1)}$.
Unfortunately this will not be possible in our setup because the stress-tensor multiplet does not appear in the bulk OPE.

Another consequence of turning on the couplings is that we expect that new infinite families of operators will enter our crossing equations.
In the bulk channel, from the intuition gained from the usual four-point function analytic bootstrap, we expect double-trace operators of the form $\phi \Box^n \phi$ with dimensions
\begin{align}
\begin{split}
  \Delta_{n}
& = d - 2 + 2n
  + \Delta_{n}^{(1)} \epsilon
  + \Delta_n^{(2)} \epsilon^2 
  + \ldots \, ,
\end{split}
\end{align}
and similarly operators $\phi \Box^n \bar \phi$ with dimensions $\tilde \Delta_n$.
In the boundary channel, we expect operators of the schematic form $\Box^{m} \partial_{\bot}^{n-2m} \phi$ so they have dimension
\begin{align}
\begin{split}
\hat \Delta_n
& = \frac{d-2}{2} + n
  + \hat \Delta_n^{(1)} \epsilon
  + \hat \Delta_n^{(2)} \epsilon^2 
  + \ldots \, .
\end{split}
\end{align}
Finally, the OPE coefficients will also get corrections as a power series in $\epsilon$, namely
\begin{align}
\begin{split}
 c_n
& = c_n^{(0)}
  + c_n^{(1)} \epsilon
  + c_n^{(2)} \epsilon^2 
  + \ldots \,,
\end{split}
\end{align}
and similarly for $d_n$, $\mu_n$ and $\rho_n$.
With the above conventions, the free theory solution when $d = 4$ is given by
\begin{align}
 d^{(0)}_0 = \mu_0^{(0)} = \rho_{0}^{(0)} = 1, \qquad
 c^{(0)}_{n\ge0} = d^{(0)}_{n\ge1} = \mu^{(0)}_{n\ge1} = \rho^{(0)}_{n\ge1} = 0 \,.
\end{align}
In what follows we derive the first order correction to the CFT data.

\subsubsection{\texorpdfstring{$\langle\phi\phi\rangle$}{<Phi Phi>} correlator}

We start by studying the two-point function $\langle \phi \phi \rangle$ , because in this case we can reuse many results from~\cite{Bissi:2018mcq}. 
We will very closely follow the notation and manipulations from this reference, and we refer the reader there for further details.
The similarity is a consequence of the $\phi\phi$ bulk channel superconformal blocks being equal to non-supersymmetric ones: $F^{\phi\phi}_\Delta = f_\Delta$.

The first step in the construction of~\cite{Bissi:2018mcq} is to divide the crossing equation in  two terms called $G$ and $H$:
\begin{align}
 \Fm^{\phi\phi}(\xi) 
 = G_{\text{blk}}(\xi) + H_{\text{blk}}(\xi) 
 = G_{\text{bdy}}(\xi) + H_{\text{bdy}}(\xi) \,.
\end{align}
In $G$ we collect the contributions that appeared at order $\epsilon^0$, but we allow them to acquire anomalous dimensions:
\begin{align}
\label{eq:GPP}
\begin{split}
 & G_{\text{blk}}(\xi) 
 = F^{\phi\phi}_{2-\epsilon + \Delta_0^{(1)} \epsilon}(\xi)
 = \frac{1}{\xi +1}
 + \frac{(\Delta_0^{(1)} -2 \Delta_\phi^{(1)}) \log \xi + \log (\xi +1)}
        {2 (\xi +1)} \epsilon 
 + \Om(\epsilon^2) \,, \\
 & G_\text{bdy}(\xi) 
 = \hat F^{\phi\phi}_{1 - \frac{\epsilon}{2} + \epsilon \hat \Delta_0^{(1)}}(\xi)
 = \frac{1}{\xi +1}
 + \frac{\log (\xi +1)-2 \hat \Delta_0^{(1)} \log \xi}{2 (\xi +1)} \epsilon
 + \Om(\epsilon^2) \,.
\end{split}
\end{align}
On the other hand, we collect in $H$ all the contributions where the anomalous dimensions do not contribute, so the blocks are evaluated at integer values of the dimensions:
\begin{align}
\label{eq:HDef}
 H_{\text{blk}}(\xi)
 = \epsilon \sum_{n=0}^\infty d_n^{(1)} F^{\phi\phi}_{2n+2}(\xi)\,, \qquad
 H_{\text{bdy}}(\xi)
 = \epsilon \sum_{n=0}^\infty \rho_n^{(1)} \hat F^{\phi\phi}_{n+1}(\xi)\,.
\end{align}
Note that an operator can contribute to both $G$ and $H$, for instance the anomalous dimension $\Delta_0^{(1)}$ of the leading bulk operator $\Om_0$ appears in $G_{\text{blk}}$, while the correction to its OPE coefficient $d_0^{(1)}$ appears in $H_{\text{blk}}$.

The key observation of~\cite{Bissi:2018mcq} was that one can eliminate $H_{\text{bdy}}$ from the crossing equation by applying the following discontinuity:
\begin{align}\label{eq:discDef}
 \text{Disc} \, f(z) = f(z e^{i\pi}) - f(z e^{-i\pi})\,, \qquad
 z \equiv \xi + \half \in (\tfrac12, \infty)\,.
\end{align}
Indeed, from~\eqref{eq:discDef} one sees that $\text{Disc}\, \hat f_n(\xi) = 0$ for integer $n$, which implies $\text{Disc}\, H_{\text{bdy}}(\xi) = 0$.
It is an easy exercise to take the discontinuity of~\eqref{eq:GPP}, and using the crossing equation we find
\begin{align}
\label{eq:DiscH}
 \text{Disc} \, H_\text{blk}(\xi) = 
 \text{Disc} \, G_\text{bdy}(\xi) - \text{Disc} \, G_\text{blk}(\xi) =
 \frac{i\pi\epsilon}{\xi} \left( 
      2 \hat \Delta_0^{(1)}
    - 2 \Delta_\phi^{(1)}
    + \Delta_0^{(1)}
 \right) \,.
\end{align}
The authors of~\cite{Bissi:2018mcq} reconstructed the full correlator by expanding~\eqref{eq:DiscH} in terms of discontinuities of bulk blocks, extracting the CFT data, and then resumming the bulk OPE expansion.
Note that since our expansion in the bulk has non-supersymmetric blocks, we can reuse their results without problems. 
In particular, comparing their equations (4.8) and (4.14) with our expression we obtain
\begin{align}
 H_\text{blk}(\xi)
 = -\frac{\epsilon  \log (\xi +1)}{2 (\xi +1)} \left( 
      2 \hat \Delta_0^{(1)}
    - 2 \Delta_\phi^{(1)}
    + \Delta_0^{(1)}
 \right)\,.
\end{align}
From this calculation we can reconstruct the full correlator $\Fm^{\phi\phi}(\xi) = G_\text{blk}(\xi) + H_\text{blk}(\xi)$ and extract CFT data to $\Om(\epsilon)$.
Before we do that, however, let us also reconstruct the $\langle \phi \bar \phi \rangle$ correlator using the same technique.

\subsubsection{\texorpdfstring{$\langle\phi\bar\phi\rangle$}{<Phi PhiBar>} correlator} 

As before, let us divide the contributions of the crossing equations into two pieces, where
\begin{align}
\begin{split}
 & G_\text{blk}(\xi)
 = F^{\phi\bar\phi}_{\text{Id}}(\xi)
 = \frac{1}{\xi}
 + \frac{(1-2\Delta_{\phi}^{(1)}) \log\xi}{2 \xi} \epsilon
 + \Om(\epsilon^2)\,, \\
 & G_\text{bdy}(\xi) 
 = \hat F^{\phi\bar\phi}_{1 - \frac{\epsilon}{2} + \epsilon \hat \Delta_0^{(1)}}(\xi)
 = \frac{1}{\xi }
 + \frac{\log \xi - 2 \hat \Delta_0^{(1)} \log (\xi +1)}{2 \xi } \epsilon
 + \Om(\epsilon^2)\,,
\end{split}
\end{align}
and the functions $H$ are the same we defined in~\eqref{eq:HDef}, replacing $(d_n, \rho_n) \to (c_n, \mu_n)$ and using the appropriate superconformal blocks for $\phi\bar\phi$.
Again, the discontinuity removes $H_\text{bdy}(\xi)$ and we are left with
\begin{align}
 \text{Disc} \, H_\text{blk}(\xi) = 
 \text{Disc} \, G_\text{bdy}(\xi) - \text{Disc} \, G_\text{blk}(\xi) =
 -\frac{2\pi i \epsilon}{\xi+1} \left( 
    \Delta_\phi^{(1)}
    - \hat \Delta_0^{(1)}
 \right) \,.
\end{align}
This can be expanded in terms of discontinuities of superconformal blocks.
In principle, we should repeat the analysis of~\cite{Bissi:2018mcq} using our superconformal blocks. However, the first term in the expansion captures the entire correlator:
\begin{align}
\label{eq:discF2}
 \text{Disc} \, F^{\phi\bar\phi}_2(\xi) 
 = -\frac{2 \pi i}{\xi +1} 
 = \frac{\text{Disc} \, H_\text{blk}(\xi)}{
  \epsilon \left( \Delta_\phi^{(1)} - \hat \Delta_0^{(1)} \right)}\,.
\end{align}
We can remove the discontinuity from this equation\footnote{The discontinuities of superblocks are schematically $\text{Disc} \, F^{\phi\bar\phi}_{2n} \sim P_{n}$, where $P_n$ are certain orthogonal polynomials. Since any function has a unique expansion in terms of $P_n$, it is safe to remove Disc from~\eqref{eq:discF2}.
 } to obtain
\begin{align}
 H_\text{blk}(\xi)
 = \epsilon \left(\Delta_\phi^{(1)} - \hat \Delta_0^{(1)} \right) F^{\phi\bar\phi}_2
 = \epsilon \left(\Delta_\phi^{(1)} - \hat \Delta_0^{(1)} \right) 
   \frac{\log(\xi+1)}{\xi}\,.
\end{align}
The full correlator is $\Fm^{\phi\bar\phi}(\xi) = G_\text{blk}(\xi) + H_\text{blk}(\xi)$.
Equation~\eqref{eq:discF2} implies that the bulk channel of $\phi\bar\phi$ contains only the identity and another block, unlike the $\phi\phi$ expansion which contained infinitely many blocks.

\subsubsection{Correlation functions and CFT data} 

The solution of crossing we have found to $\Om(\epsilon)$ has three free parameters.
However, as discussed below equation~\eqref{eq:crossingEps}, the boundary OPE coefficients in the two channels should be equal up to a sign $\rho_n = \pm \mu_n$.
Expanding $\Fm^{\phi\phi}(\xi)$ and $\Fm^{\phi\bar\phi}(\xi)$ in boundary superblocks and comparing the expansions we find one last constraint:
\begin{align}
\label{eq:anomDimD01}
 \Delta_{0}^{(1)} =
 2 \left( (s+1) \Delta_\phi^{(1)} - \hat \Delta_0^{(1)} \right)\,, 
 \qquad s = \pm\,.
\end{align}
Hence, our solution depends on the anomalous dimension $\Delta_\phi^{(1)}$ of the external chiral operator, the anomalous dimension of the leading boundary operator $\hat \Delta_0^{(1)}$, and a choice of signs $s = \pm$.
Using this relation, the one-loop correlation functions take a very simple form\footnote{We can also write $\langle\phi(x) \bar \phi(y) \rangle = (x-y)^{-2\Delta_\phi} (\xi+1)^{\epsilon(\Delta_\phi^{(1)} - \hat\Delta_0^{(1)})}$ and similarly for $\langle\phi\phi\rangle$.
This is very similar to the non-supersymmetric case, see equation (2.32) of~\cite{Prochazka:2019fah}. We thank A.~S{\"o}derberg for pointing this out.}
\begin{align}
\label{eq:corr-bootstrap}
\begin{split}
 & \Fm^{\phi\bar\phi}(\xi)
 = \frac{1}{\xi}
 + \frac{
    \left( 1 - 2 \Delta_\phi^{(1)} \right) \log \xi
    + 2 \left( \Delta_\phi^{(1)} - \hat\Delta_0^{(1)} \right) \log (\xi +1)
    }{2 \xi } \epsilon
 + \Om(\epsilon^2)\,, \\
 & \Fm^{\phi\phi}(\xi) 
 = \frac{1}{\xi +1}
 +\frac{
    \left( 1 - 2 s \Delta_\phi^{(1)} \right) \log (\xi +1) 
    + 2 \left( s \Delta_\phi^{(1)} - \hat\Delta_0^{(1)} \right) \log \xi }
    {2 (\xi +1)} \epsilon
 + \Om(\epsilon^2)\,.
\end{split}
\end{align}
From the correlation functions we can extract the CFT data at one-loop:\footnote{
Our solution of crossing splits naturally into a pice involving only boundary blocks with $n=0$ and a piece that includes all $n \ge 1$.
This resembles the four-point analytic bootstrap where our $n$ plays the role of the bulk spin $\ell$.
In particular, our $n=0$ solution corresponds to the solutions with finite support in spin found in~\cite{Alday:2016jfr}.
We thank F. Alday for pointing this out.}
\begin{align}
\label{eq:cft-data}
\begin{split}
 & c_0^{(1)}
 = \Delta_\phi^{(1)} - \hat \Delta_0^{(1)}, \qquad
 c_{n\ge1}^{(1)} = 0, \qquad 
 d_0^{(1)} = 0, \qquad
 \mu_0^{(1)} = \rho_0^{(1)} = 0\,, \\[0.5em]
 & \mu_n^{(1)} 
 = s (-1)^{n} \rho_n^{(1)}
 = s (-1)^{n} d_n^{(1)}
 = \frac{(n-1)!}{2^{n-1}(2n-1)!!} \Delta_\phi^{(1)}, 
 \qquad n \ge 1\,.
\end{split}
\end{align}
Although we lack a conclusive proof, we believe it is very likely that the unfixed sign is always $s = +1$.
One argument is that the correlators~\eqref{eq:corr-bootstrap} are related to each other under $\xi \leftrightarrow \xi + 1$, provided $s = +1$.
Another argument is that only for $s = +1$ the signs of the coefficients in the BOE are alternating, namely $(\rho_0, \rho_1, \rho_2, \ldots) = (\mu_0, -\mu_1, \mu_2, \ldots)$, and otherwise they are alternating only for $n \ge 1$.
Finally, we will do an explicit perturbative calculation for a specific model in the next section and we will find again that $s = +1$.

An interesting feature of the CFT data~\eqref{eq:cft-data} is that the bulk and boundary OPE coefficients are identical for the two-point function $\langle \phi \phi \rangle$ .
This is a very non-trivial relation, since $\rho_n = b_{\phi\Om_n}^2$, but $d_n = a_{\Om_n} \lambda_{\phi\phi\Om_n}$. 
It would be interesting to see if this is just a coincidence of the order $\Om(\epsilon)$ result, or if it actually persists at higher orders in perturbation theory.

\subsubsection{Going to order \texorpdfstring{$\epsilon^2$}{epsilon\^2}} 

From the structure of the order $\epsilon$ CFT data~\eqref{eq:cft-data}, there is hope that one can push the bootstrap analysis to order $\epsilon^2$.
Indeed, only two blocks contribute at order $\epsilon$ in the $\phi\bar\phi$ bulk channel. 
We expect infinitely many operators at order $\epsilon^2$, but the majority of them will contribute as conformal blocks of even dimension $F_{2n+2}(\xi)$.
One can construct a discontinuity, different than~\eqref{eq:discDef}, that kills bulk blocks $\widetilde{\text{Disc}} \, F_{2n + 2} = 0$, see~\cite{Mazac:2018biw}.
From here there are several possible directions one can pursue:
\begin{itemize}
 \item Following the ideas of the present section and~\cite{Bissi:2018mcq}, one can calculate $\widetilde{\text{Disc}} \, H_{\text{bdy}} = \widetilde{\text{Disc}} \, G_{\text{blk}}- \widetilde{\text{Disc}} \, G_{\text{bdy}}$. One should now expand $\widetilde{\text{Disc}} \, H_{\text{bdy}}$ in terms of discontinuities of boundary blocks to extract the relevant CFT data. However, at this order in $\epsilon$, the discontinuities of the blocks cannot be easily rewritten in terms orthogonal polynomials, and it is not clear how to proceed.
 \item The authors of~\cite{Mazac:2018biw} studied an inversion formula that would reconstruct the boundary data from the two discontinuities $\text{Disc} \, \Fm$ and $\widetilde{\text{Disc}} \, \Fm$ of a correlator.
 Unfortunately, they were unable to determine its precise form for the case of interest here, and even if the relevant inversion formula is found, calculating $\text{Disc} \, \Fm^{\phi\bar\phi}$ in our setup would be challenging.
 \item Finally, one can make an ansatz for the full correlator based on trascendentality and demand consistency with the above discontinuities to fix coefficients.
 With this approach it is possible to rederive the order $\epsilon^2$ correlator of the Wilson-Fischer fixed point calculated in \cite{Bissi:2018mcq}. In our supersymmetric setup, we have found a consistent solution to crossing at order $\epsilon^2$ that depends on a number of free parameters.
 However, it is not clear to us yet whether this correlator is physical or whether it is part of a more general solution of crossing yet to be found.
\end{itemize}

\section{Wess-Zumino model with a boundary}
\label{sec:wess-zumino}

In this section we study the Wess-Zumino (WZ) model with a cubic superpotential in the presence of half-BPS boundary conditions.
The WZ model has a stable fixed point in $4-\epsilon$ dimensions, which has been studied in the context of emergent supersymmetry~\cite{Lee:2006if,Fei:2016sgs}.
The two-loop calculation of~\cite{Townsend:1979ha} showed that supersymmetry is preserved perturbatively, provided the gamma matrix algebra is evaluated in $d=4$, but using a $4-\epsilon$ dimensional spacetime otherwise.
Here we adopt the same regularization procedure, which is reminiscent of the way we obtained the blocks in section~\ref{sec:4-eps}, using a superconformal algebra with $4d$ spinor representations, but allowing arbitrary $d \le 4$ spacetime dimensions.
Furthermore, we assume that the boundary is exactly codimension-one for any $d$.

\subsection{Action and boundary conditions}

Our model consists of a single chiral multiplet interacting with a cubic superpotential,
so the degrees of freedom are the real and imaginary parts of $\phi = \phi_1 + i \phi_2$, a four-component Majorana fermion $\Psi$, and the real and imaginary parts of the auxiliary fields $F = F_1 + i F_2$.
The action is obtained by integrating the Lagrangian density over a half-space, with parallel coordinates $\mathbf x \in \mathbb{R}^{d-1}$ and perpendicular coordinate $y \in \mathbb R^+$:\footnote{
We work in Euclidean signature with $\{ \gamma^\mu, \gamma^\nu \} = 2 \delta^{\mu\nu}$ and $\gamma_5  = \gamma^1 \gamma^2 \gamma^3 \gamma^4$. 
The Majorana reality condition is $\Psi^T \Cm = \bar \Psi$, where the charge conjugation matrix satisfies $\gamma^\mu = - \Cm^{-1} (\gamma^\mu)^T \Cm$.
} 
\begin{align}
\label{eq:off-shell-action}
\begin{split}
 S_{blk} = \int d^{d-1} \mathbf x \, dy \bigg(
  &
      \frac12 (\partial_\mu \phi_1)^2
    + \frac12 (\partial_\mu \phi_2)^2
    + \frac{1}{2} \bar \Psi \gamma^\mu \partial_\mu \Psi
    - \frac{1}{2} F_1^2
    - \frac{1}{2} F_2^2 \\
  & - \frac{\lambda}{2\sqrt{2}} \Big(
      F_1 (\phi_1^2 - \phi_2^2) 
    + 2 F_2 \phi_1 \phi_2
    - \bar \Psi (\phi_1 + i \gamma_5 \phi_2) \Psi \Big)
 \bigg)\, .
\end{split}
\end{align}
In order to compute Feynman diagrams, it will be simpler to integrate out the auxiliary fields $F_i$, producing the following interaction vertices:
\begin{align}
\label{eq:int-vtx}
\begin{split}
 S_{int} = \int d^{d-1} \mathbf x \, dy \bigg(
      \frac{1}{16} \lambda^2 \left( \phi_1^2 + \phi_2^2 \right)^2
    + \frac{\lambda}{2\sqrt{2}} \bar \Psi (\phi_1 + i \gamma_5 \phi_2) \Psi
 \bigg) \,.
\end{split}
\end{align}
However, it is easier to work with the off-shell action to study how the boundary breaks supersymmetry.
The supersymmetry transformations are parametrized by a Majorana spinor $\epsilon$ and they are well known:
\begin{align}
\label{eq:susy-trans}
\begin{split}
 & \delta \phi_1 = - \bar \epsilon \Psi, \\
 & \delta \phi_2 = i \bar \epsilon \gamma_5 \Psi\,,
\end{split}
\begin{split}
 & \delta \Psi = \left(- \slashed \partial \phi_1
                  - i \gamma_5 \slashed \partial \phi_2
                  - F_1
                  + i \gamma_5 F_2 \right) \epsilon\,, \\
 & \delta \bar \Psi = \bar \epsilon \left(\slashed \partial \phi_1
                  - i \gamma_5 \slashed \partial \phi_2
                  - F_1
                  + i \gamma_5 F_2 \right)\,,
\end{split} \qquad
\begin{split}
 & \delta F_1 = - \bar \epsilon \slashed \partial \Psi, \\
 & \delta F_2 = i \bar \epsilon \gamma_5 \slashed \partial \Psi\,.
\end{split}
\end{align}
If we integrated the Lagrangian~\eqref{eq:off-shell-action} over $\mathbb{R}^d$, the supersymmetry transformations~\eqref{eq:susy-trans} would be an exact symmetry of the action. However, the situation is more complicated in the presence of the boundary.
On the one hand, we know that not all supersymmetries can be preserved, because that would imply that translations orthogonal to the boundary are also preserved.
We can preserve at most half of the supersymmetry, namely the transformations generated by spinors satisfying~\cite{Herzog:2018lqz}
\begin{align}
 \Pi_+ \epsilon = \epsilon 
 \quad \Leftrightarrow \quad
 \Pi_- \epsilon = 0, \qquad
 \Pi_{\pm} \equiv \frac{1}{2} \left( 1 \pm i \gamma_5 \gamma^n \right)\,.
\end{align}
On the other hand, to check invariance under supersymmetry of~\eqref{eq:off-shell-action}, we have to integrate by parts, which generates extra boundary terms.
Supersymmetry will only be preserved for an action containing extra boundary degrees of freedom $S = S_{blk} + S_{bdy}$, provided we choose $S_{bdy}$ to cancel the terms generated by the supersymmetry variation of $S_{blk}$.
A systematic study of all possible boundary actions for a generic $4d$ $\Nm = 1$ theory appeared in~\cite{Bilal:2011gp}, and we can easily translate their results to our conventions.
For the purposes of this section, we will pick the minimal boundary action that preserves supersymmetry, although more general options would be possible:
\begin{align}
 S_{bdy} = \int d^{d-1} \mathbf x \left( 
      \frac12 \Big( 
         \phi_1 \partial_n \phi_1
       + \phi_2 \partial_n \phi_2
       + \phi_1 F_2
       + \phi_2 F_1 \Big)
    - \frac{\lambda}{2\sqrt{2}} 
      \Big( \tfrac{1}{3} \phi_2^3 - \phi_1^2 \phi_2 \Big)
 \right)\,.
\end{align}
It is an easy but tedious exercise to check that the combination of bulk and boundary actions indeed preserves half of the original supersymmetries.

Next we address the problem of determining the boundary conditions of our fields.
Demanding that the Euler-Lagrange variation of the total action vanishes produces a bulk term which is zero, provided that the fields satisfy the equations of motion (EOM).
However, we also get terms localized in the boundary
\begin{align}
\label{eq:el-var}
\begin{split}
  \delta(S_{bulk} & + S_{bdy})
  = \int d^{d-1} \mathbf x \, dy \left( \, \mathrm{EOM} \, \right)
  + \int d^{d-1} \mathbf x \bigg( 
      \frac12 \phi_1 \, \delta(F_2 + \partial_n \phi_1) \\
&   + \frac12 \phi_2 \, \delta(F_1 + \partial_n \phi_2)
    - \frac12 \delta \phi_1 (F_2 + \partial_n \phi_1) 
    - \frac12 \delta \phi_2 (F_1 + \partial_n \phi_2) 
    + \frac{1}{2} \delta \bar \Psi \gamma^n \Psi
  \bigg)\,,
\end{split}
\end{align}
and the boundary conditions must be chosen such that they are zero.
Moreover, one must check that the boundary conditions are closed under the supersymmetry transformations ~\eqref{eq:susy-trans}.
In~\cite{Bilal:2011gp} it was shown that there is only one possible supersymmetric boundary condition, up to $R$-symmetry redefinitions $\phi \to e^{i\delta} \phi$.
In conventions that match the bootstrap analysis of section~\ref{sec:4-eps} this boundary condition is
\begin{align}
\label{eq:WZ-bc}
 \partial_n \phi_1 = - F_2 = \frac{\lambda}{\sqrt{2}} \phi_1 \phi_2\,, \qquad
 \phi_2 = 0, \qquad
 \Pi_- \Psi = 0\,.
\end{align}
In equations~\eqref{eq:el-var} and~\eqref{eq:WZ-bc} we used the bulk equations of motion that relate $F \sim \lambda \phi^2$.
Since we will work in perturbation theory, the free propagators are obtained for $\lambda = 0$, where $\phi_1$ satisfies Neumann boundary conditions $\partial_n \phi_1 = 0$.
As pointed out in~\cite{Diehl:2020rfx}, these boundary conditions are a good description near the free theory, but are not meant to describe the boundary condition of the fields at the interacting fixed point.

\subsection{Using susceptibility}

The calculation of correlation functions in the presence of boundaries using Feynman diagrams is typically challenging.
An important observation that dates back to the work of McAvity and Osborn~\cite{McAvity:1995zd,McAvity:1995bh} is that the calculations simplify in terms of susceptibilities, defined as
\begin{align}
\label{eq:sus-def}
 \chi_{\Om_1\Om_2}(y, y') 
 = \int d^{d-1} \mathbf x \, \langle \Om_1(\mathbf x, y) \Om_2(0, y') \rangle \,.
\end{align}
Crucially, this integral transform is invertible and one can recover the two-point function in terms of the susceptibility. 
This idea has been recently used to compute the one-loop two-point function of the order parameter in the extraordinary phase transition of the $O(N)$ model~\cite{Shpot:2019iwk,Dey:2020lwp}.
One can also apply it to the $O(N)$ model in the large-$N$ limit, see~\cite{Herzog:2020lel} for the three-dimensional case with a $\phi^6$ potential.

In susceptibility space the role of the cross ratio $\xi$ is played by a new object $\zeta$, which is defined as follows:
\begin{align}
 \zeta 
 = \frac{\min(y, y')}{\max(y, y')}\,.
\end{align}
The importance of $\zeta$ was noted in~\cite{Dey:2020lwp}, where they observed that the contribution of a single conformal block in the boundary expansion is proportional to $\zeta^{\hat\Delta-\frac{d-1}{2}}$.
This allows one to extract the boundary CFT data directly from the susceptibility without the need to reexpress everything in terms of the correlation function $F(\xi)$.
To be more precise, the susceptibility can be expanded as
\begin{align}
 \chi_{\Om\Om}(y, y') 
 = (4yy')^{\frac{d-1}{2} - \Delta_\Om}
   \sum_{\hat\Om} \mu_{\hat\Om} \pi^{\frac{d-1}{2}} 
   \frac{\Gamma\left( \hat\Delta - \frac{d-1}{2}\right)}{\Gamma(\hat\Delta)}
   (4 \zeta)^{\hat\Delta-\frac{d-1}{2}}\,,
\end{align}
where $\Delta_\Om$ is the dimension of the external operator, $\mu_{\hat\Om}$ is the boundary OPE coefficient and $\hat\Delta$ is the dimension of the exchanged operator.

Even though the bootstrap analysis used the chiral field and its complex conjugate, for the purposes of the current section it is more convenient to work with its real and imaginary parts $\phi = \phi_1 + i \phi_2$.
The susceptibilities of the two descriptions are related by
\begin{align}
\begin{split}
 & \chi^+(y, y') 
 \equiv \chi_{\phi_1\phi_1}(y,y') 
 = \frac{1 + \frac{\epsilon}{2}(\gamma + \log \pi)}{4\pi^2} 
   \left( \chi_{\phi\bar\phi}(y,y') + \chi_{\phi\phi}(y,y') \right)\,,  \\
 & \chi^-(y, y') 
 \equiv \chi_{\phi_2\phi_2}(y,y') 
 = \frac{1 + \frac{\epsilon}{2}(\gamma + \log \pi)}{4\pi^2}
   \left( \chi_{\phi\bar\phi}(y,y') - \chi_{\phi\phi}(y,y') \right)\,,
\end{split}
\end{align}
where the prefactor translates from the natural normalization in the bootstrap calculation to the natural normalization using Lagrangians.
It is an easy exercise to check that our prediction for the order $\epsilon$ correlator~\eqref{eq:corr-bootstrap} leads to
\begin{align}
\label{eq:boot-pred}
\begin{split}
 \chi^+(y, y')
 & = \frac{-1}{2\sqrt{\zeta}} (4yy')^{\frac{1}{2}-\Delta_\phi^{(1)} \epsilon } 
   \bigg[
        1
    + 2 \hat \Delta_{0}^{(1)} \epsilon
    + \hat\Delta_{0}^{(1)} \epsilon \log \zeta \\
 &  \qquad \quad \quad \quad
    - \Delta_\phi^{(1)} \epsilon 
      \big( (1+\zeta) \log (1+\zeta) + (1-\zeta ) \log (1-\zeta ) \big) 
    + \Om(\epsilon^2)
 \bigg]\,, \\
 \chi^-(y, y')
 & = \frac{\sqrt{\zeta}}{2} (4yy')^{\frac{1}{2}-\Delta_\phi^{(1)} \epsilon } 
   \bigg[
     1
    + 2 \left( 2\Delta_\phi^{(1)} - \hat \Delta_{0}^{(1)} \right) \epsilon
    + \hat\Delta_{0}^{(1)} \epsilon \log \zeta \\
 &  \qquad \quad \quad \quad
    - \Delta_\phi^{(1)} \epsilon 
      \left( 
          \frac{(1+\zeta)}{\zeta} \log (1+\zeta) 
        - \frac{(1-\zeta)}{\zeta} \log (1-\zeta ) \right) 
    + \Om(\epsilon^2)
 \bigg]\,.
\end{split}
\end{align}
In the rest of this section we will check that perturbation theory gives a result consistent with this prediction, and we will find the explicit values of $\Delta_\phi^{(1)}$ and $\hat\Delta_{0}^{(1)}$ for the Wess-Zumino model.

\subsection{Susceptibility at one-loop}

\subsubsection{Tree level}

To compute the scalar propagators we have to solve the Klein-Gordon equation in position space.
It is well known that in the presence of a boundary one has to add a ``mirror'' term to the propagator to satisfy the correct boundary conditions at $y = 0$.
Since $\phi_1$/$\phi_2$ satisfy Neumann/Dirichlet boundary conditions we have:
\begin{align}
\label{eq:scalar-props}
\begin{split}
  \langle \phi_1(x) \phi_1(x') \rangle_0
& = \kappa_s \left(
    \frac{1}{|x - x'|^{d-2}}
  + \frac{1}{|\bar x - x'|^{d-2}} \right), \\
  \langle \phi_2(x) \phi_2(x') \rangle_0
& = \kappa_s \left(
    \frac{1}{|x - x'|^{d-2}}
  - \frac{1}{|\bar x - x'|^{d-2}} \right)\,.
\end{split}
\end{align}
Here $\langle \ldots \rangle_0$ indicates the two-point functions are evaluated in the free theory. For each $x = (\mathbf x, y)$ we defined the mirror point $\bar x = (\mathbf x, -y)$, and the overall normalization is $\kappa_s = \frac{\Gamma(\frac{d}{2})}{(d-2)2\pi^{d/2}}$.
We will be mostly interested in the susceptibilities, which can be readily obtained from~\eqref{eq:sus-def} and~\eqref{eq:scalar-props}:
\begin{align}
\label{eq:tree-sus}
\begin{split}
 & \chi^+_0(y, y') = \chi_{\langle\phi_1\phi_1\rangle_0}(y, y') = - \max(y, y')\,, \\
 & \chi^-_0(y, y') = \chi_{\langle\phi_2\phi_2\rangle_0}(y, y') = + \min(y, y')\,.
\end{split}
\end{align}
Similarly, solving the Dirac equation and adding a ``mirror'' term dictated by the boundary conditions one gets~\cite{Herzog:2018lqz}
\begin{align}
  \langle \Psi(x) \bar \Psi(x') \rangle_0
& = \kappa_f \left( 
    \frac{\gamma \cdot (x-x')}{|x - x'|^{d}}
  + i \gamma_5 \gamma^n \frac{\gamma \cdot (\bar x - x')}{|\bar x - x'|^d} \right)\,, \qquad
  \kappa_f = \frac{\Gamma(\frac d2)}{2\pi^{d/2}}\,.
\end{align}
It is not hard to check that the fermion propagator satisfies the correct boundary conditions:
\begin{align}
 \Pi_- \langle \Psi(\mathbf x, 0) \bar\Psi(\mathbf x', y') \rangle_0
 = \langle \Psi(\mathbf x, y) \bar\Psi(\mathbf x', 0) \rangle_0 \, \Pi_-
 = 0\,.
\end{align}

\subsubsection{Tadpole diagram}

First we consider the quartic interaction terms in~\eqref{eq:int-vtx} and we use them to form loop diagrams with either $\phi_1$ or $\phi_2$ running in the loop.
These diagrams would vanish if the boundary was not present, or equivalently if we studied physics far away from the boundary. 
As a result, we expect them to be finite in the limit $\epsilon \to 0$.
Taking symmetry factors into account the total contribution is
\begin{align}
 \chi^{\pm}(y, y')|_{\text{tadpole}}
 =
 \diagramEnvelope{\begin{tikzpicture}[anchor=base,baseline]
	\node (x1) at (-1,0) [] {};
	\node (y) at (0,0) [vertex] {};
	\node (x2) at (1,0) [] {};
    \node (loop1) at (-.01,.7) [point] {};
    \node (loop2) at (.01,.7) [point] {};
	\draw [sus] (x1) -- (y);
	\draw [sus] (y) -- (x2);
	\draw [sus] (y) to[out=45,in=0] (loop1);
	\draw [sus] (y) to[out=135,in=180] (loop2);
\end{tikzpicture}}
 = \mp 2^{-3+\epsilon} \lambda^2 \kappa_s I^{\pm}_b(y, y')\,.
\end{align}
The propagator that runs in the loop is defined as the finite part of $\langle \phi_i(\mathbf x, y) \phi_i(\mathbf x', y') \rangle_0$ when $\mathbf x' \to \mathbf x$, and can be obtained from~\eqref{eq:scalar-props}.
With this prescription, the Feynman integrals we must compute are~\cite{Shpot:2019iwk}
\begin{align}
\label{eq:tad-sus}
\begin{split}
 & I^{+}_b(y, y')
 = \int_0^\infty dz \, \chi^+_0(y, z) \, z^{-2+\epsilon} \, \chi^+_0(z, y')
 = \frac{y^\epsilon y'}{\epsilon-1}
 - \frac{y' (y^\epsilon - y^{\prime\epsilon})}{\epsilon}
 - \frac{y^{\prime\epsilon+1}}{\epsilon+1} \, , \\
 & I^{-}_b(y, y')
 = \int_0^\infty dz \, \chi^-_0(y, z) \, z^{-2+\epsilon} \, \chi^-_0(z, y')
 = 
 - \frac{y y^{\prime\epsilon}}{\epsilon-1}
 - \frac{y (y^\epsilon - y^{\prime\epsilon})}{\epsilon}
 + \frac{y^{\epsilon+1}}{\epsilon+1} \, .
\end{split}
\end{align}
For simplicity we assumed here and in the rest of the section that $y < y'$, but one can obtain the integral for $y>y'$ replacing $y \leftrightarrow y'$.

\subsubsection{Fermion bubble}

Similarly, we can use the Yukawa interactions in~\eqref{eq:int-vtx} to form diagrams with fermions running in the loop.
If the boundary was not present, these diagrams would be UV divergent and would contribute to the renormalization of $\phi$.
Since the boundary does not change the UV behaviour of the theory, we expect a divergence as $\epsilon \to 0$ which is canceled by the counterterm $\delta_\phi$:
\begin{align}
\begin{split}
 \chi^{\pm}(y, y')|_{\text{bubble}}
 & =
 \diagramEnvelope{\begin{tikzpicture}[anchor=base,baseline]
	\node (x1) at (-1.5,0) [] {};
	\node (y1) at (-.5,0) [vertex] {};
	\node (y2) at (.5,0) [vertex] {};
	\node (x2) at (1.5,0) [] {};
	\draw [sus] (x1) -- (y1);
	\draw [prop] (y1) to[out=80,in=100] (y2);
	\draw [prop] (y1) to[out=-80,in=-100] (y2);
	\draw [sus] (y2) -- (x2);
\end{tikzpicture}}
+
 \diagramEnvelope{\begin{tikzpicture}[anchor=base,baseline]
	\node (x1) at (-1,0) [] {};
	\node[star,star points=5,star point ratio=2, scale=0.5, fill=black] (y1) at (0,0) [draw] {};
	\node (x2) at (1,0) [] {};
	\draw [sus] (x1) -- (y1);
	\draw [sus] (y1) -- (x2);
\end{tikzpicture}} \\
& = \lambda^2 \kappa_f^2 I_f^\pm(y, y')
- \delta_\phi \chi^{\pm}_0(y, y') .
\end{split}
\end{align}
Using the identities
\begin{align}
\label{eq:ferm-traces}
\begin{split}
 & \tr \Big[ 
       \langle \Psi(x) \bar \Psi(x') \rangle_0
       \langle \Psi(x') \bar \Psi(x) \rangle_0 \Big]
 = - 4 \kappa_f^2 \left( 
   \frac{1}{|x-x'|^{2(d-1)}} + \frac{1}{|\bar x-x'|^{2(d-1)}} \right)\,, \\
 & \tr \Big[ 
       \langle \Psi(x) \bar \Psi(x') \rangle_0 \gamma_5
       \langle \Psi(x') \bar \Psi(x) \rangle_0 \gamma_5 \Big]
 = 4 \kappa_f^2 \left( 
   \frac{1}{|x-x'|^{2(d-1)}} - \frac{1}{|\bar x-x'|^{2(d-1)}} \right)\,,
\end{split}
\end{align}
we see that the Feynman integral is
\begin{align}
 I_f^\pm(y, y')
 & = \int_0^\infty dz \int_0^\infty dz' \chi^{\pm}(y, z) b^{\pm}(z, z') \chi^{\pm}(z', y')\,,
\end{align}
where we have defined
\begin{align}
\begin{split}
  b^{\pm}(z, z') 
& = \int d^{d-1} \mathbf r \left( 
         \frac{1}{\big( \mathbf r^2 + (z - z')^2 \big)^{d-1}}
     \pm \frac{1}{\big( \mathbf r^2 + (z + z')^2 \big)^{d-1}}
  \right) \\
& = \frac{2^{2-d} \pi ^{d/2}}{\Gamma \left(\frac{d}{2}\right)}
    \left( |z-z'|^{-3+\epsilon} \pm |z+z'|^{-3+\epsilon} \right)\,.
\end{split}
\end{align}
We will evaluate this integral with a trick that has been used in the literature in similar situations~\cite{PhysRevB.50.10009,Shpot:2019iwk,Dey:2020lwp}.
The idea is to split the integration region between $z>z'$ and $z<z'$. By changing variables to $Z = \frac{z}{z'}$ and $Z = \frac{z'}{z}$, one can carry out the first integration in terms of $I_b^{\pm}$ defined in the previous section. 
The result is:
\begin{align}
\label{eq:last-int}
   I_f^\pm(y, y')
 & = y^{\epsilon+1} \Bigg[
     \int_1^\infty d Z I^{\pm}_b(1, Z / \zeta) b^\pm(1, Z) Z^{-\epsilon}
   + \int_1^{1/\zeta} dZ I^{\pm}_b(1, (Z\zeta)^{-1}) b^\pm(1, Z) Z \nonumber \\
 & \quad
   + \zeta^{-1-\epsilon} 
     \int_{1/\zeta}^\infty dZ 
     I^{\pm}_b(1, \zeta Z) b^\pm(1, Z) Z^{-\epsilon}
   \Bigg]\,.
\end{align}
Remember that we are assuming $y<y'$, such that $\zeta = y / y'$.
Finally, all terms in~\eqref{eq:last-int} can be integrated using \texttt{Mathematica}\footnote{The only exception are integrals of the form $\int_1^\infty dZ (Z-1)^a$, but they are zero in dimensional regularization.}.
The result for general $\epsilon$ is not particularly illuminating and will not be needed later, instead we focus on the result in the limit $\epsilon \to 0$.
First, the divergent piece is canceled in $\overline{\text{MS}}$ with the following counterterm:
\begin{align}
 \delta_\phi 
 = -\frac{\lambda^2}{(4\pi)^2} 
   \left( \frac{1}{\epsilon} + \frac{1}{2} (\gamma +\log \pi) +  1 \right)\,.
\end{align}
The total diagram is now finite:
\begin{align}
\label{eq:bub-sus}
\begin{split}
 \chi^+(y, y')|_{\text{bubble}}
 & = \frac{\lambda ^2}{64 \pi ^2} \frac{\sqrt{4yy'}}{\sqrt{\zeta}}
   \Big(
       \log (4yy')
     - \log \zeta
     - 2 \\
 &   \qquad \qquad \qquad \quad
     + (1 + \zeta ) \log (1 + \zeta)
     + (1 - \zeta ) \log (1 - \zeta)
   \Big) + \Om(\epsilon)\,, \\
 \chi^-(y, y')|_{\text{bubble}}
 & = \frac{-\lambda ^2}{64 \pi ^2} \sqrt{4yy' \zeta}
   \Big(
       \log (4yy') 
     - \log \zeta 
     - 2 \\
 &   \qquad \qquad \qquad \quad
     + \frac{(1 + \zeta)}{\zeta} \log (1 + \zeta)
     - \frac{(1 - \zeta)}{\zeta} \log (1 - \zeta)
   \Big) + \Om(\epsilon)\,.
\end{split}
\end{align}

\subsubsection{Final result}
\label{sec:fin-res}

We can obtain the full susceptibility at order $\epsilon$ by combining the tree-level result~\eqref{eq:tree-sus}, the tadpole diagram~\eqref{eq:tad-sus}, and the fermion bubble~\eqref{eq:bub-sus}.
We should evaluate the sum at the fixed point coupling $\lambda_*^2 = \frac{16 \pi^2}{3} \epsilon$, and keep only terms up to order $\epsilon$.
The result is perfectly consistent with the bootstrap prediction~\eqref{eq:boot-pred}, and we identify 
\begin{align}
\label{eq:anom-dim-fin}
 \Delta_\phi^{(1)} = \frac{1}{6}\,, \qquad
 \hat \Delta_{0}^{(1)} = 0\,, \qquad
 s = +1\,.
\end{align}
The anomalous dimension of $\phi$ in the Wess-Zumino model is well known in the literature. 
One can obtain it by demanding that the superpotential has $R$-charge $R(W) = 3 \mathit{r}_\phi = 2$, so we find that $\mathit{r}_\phi = 2/3$. Using the relation between the $R$-charge and conformal dimension we find $\Delta_\phi = \frac{d-1}{3}$, in perfect agreement with~\eqref{eq:anom-dim-fin}.
From this argument it is clear that $\Delta_\phi$ is one-loop exact.

An interesting prediction of our calculation is the anomalous dimension of the leading bulk operator in the OPE $\phi \times \phi \sim \Om_0 + \ldots$. 
We calculated this for a general model in~\eqref{eq:anomDimD01}, and for the Wess-Zumino case we get
\begin{align}
 \Delta_0^{(1)} = \frac{2}{3}
 \qquad \Rightarrow \qquad
 \Delta_0 = \frac{d+2}{3} = d - 2\Delta_\phi\,.
\end{align}
Recalling the selection rules of section~\ref{sec:bulk-4eps}, we see that the exchanged operator is of the form $\Om_0 \sim (Q^+)^2 \bar \Psi$ where $\bar\Psi$ is an antichiral primary operator.
Indeed, the numerical bootstrap applied to the Wess-Zumino model in~\cite{Bobev:2015vsa,Bobev:2015jxa} also provides strong evidence that the leading operator in the $\phi \times \phi$ OPE is of this form.
The agreement of our results with the predictions from~\cite{Bobev:2015vsa,Bobev:2015jxa} provides a non-trivial sanity check of our perturbative calculation.
It would be interesting to consider other particular models, for instance with extra boundary interactions or a more complicated bulk, and see whether the anomalous dimension of the defect operator changes.
We hope to come back to this question in the future.

\section{Conclusions}
\label{sec:conclusions}

In this work we studied supersymmetric boundaries for $3d$ $\Nm=2$ superconformal theories. There are two possible choices characterized by $2d$  $\Nm=(0,2)$ and $\Nm=(1,1)$ boundary algebras respectively. After performing a careful superspace analysis of correlators involving chiral fields, we observed in section~\ref{sec:4-eps} that the $\Nm=(1,1)$ choice can be analytically continued in the spacetime dimension. This allowed us to compute superconformal blocks across dimensions and opened the door for the $\epsilon$-expansion bootstrap in our supersymmetric setup. We proved uniqueness of the first two orders in $\epsilon$, and confirmed our general prediction for one specific model using perturbation theory. We used standard perturbative and bootstrap techniques in this analysis, but one could also try using alternative approaches, such as Mellin space~\cite{Kaviraj:2018tfd} or the equations of motion method of~\cite{Giombi:2020rmc}.

An interesting follow up to our BCFT analysis is to consider higher codimension defects. 
The algebraic approach to calculate superblocks of section~\ref{sec:4-eps} is applicable to higher codimension, where the spacetime and defect dimensions are allowed to change while the codimension is kept fixed. 
In particular, the codimension-two blocks calculated in section~\ref{sec:codimension-two} are applicable to known examples, such as the Wess-Zumino model in the presence of twist defects.
In the same spirit of the present work, one can use the $\epsilon$-expansion to study two-point functions of local operators.
In principle one can do explicit perturbative calculations as in~\cite{Gaiotto:2013nva}, however it is perhaps simpler to set up a bootstrap problem and attempt to solve it using the technology of inversion formulas~\cite{Lemos:2017vnx,Liendo:2019jpu}. 
One could also concentrate exclusively on three dimensions and apply the numerical bootstrap on the line, analogous to what was done for the twist defect in the $3d$ Ising model \cite{Gaiotto:2013nva} (the $1d$ bootstrap for superconformal line defects has been studied in~\cite{Liendo:2018ukf,Gimenez-Grau:2019hez,Bianchi:2020hsz}).

A longer term goal is to include multi-point correlators in the analysis; this is a program that has been underexplored even in the bosonic case, although significant progress can be made using Calogero-Sutherland technology \cite{Buric:2020zea}. Finally, the study of free theories in the presence of interacting defects has gotten some attention recently; in particular the results of \cite{Behan:2020nsf} suggest the existence of a new conformal boundary condition for the free scalar field.
It would be interesting to repeat their analysis in our supersymmetric setup, either for boundaries or higher codimension defects.

\acknowledgments
We are particularly grateful to E.~Lauria for many discussions and collaboration during several stages of this project.
We thank C.~Beem, A.~Bissi, I.~Buric, Z.~Liu, J.~Rong, V.~Schomerus, and A.~S{\"o}derberg for useful discussions and comments. 
We also thank the Simons Collaboration on the Non-perturbative Bootstrap for many stimulating activities. 
This work is supported by the DFG through the Emmy Noether research group ``The Conformal Bootstrap Program'' project number 400570283.   

\appendix

\section{Details on three-dimensional boundaries}
\label{app:3d}

\subsection{Conventions}
\label{sec:conventions}

In section~\ref{sec:3d} we work in Lorenzian signature with mostly plus metric $\eta_{\mu\nu} = \text{diag}(-1, +1, +1)$.
The gamma matrices are defined in terms of the identity matrix $\mathbf{1}$ and Pauli matrices $\sigma^i$ as
\begin{equation}
(\gamma^{\mu})_{\alpha \beta} \equiv \left(- \mathbf{1}_{\alpha \beta}, (\sigma^3)_{\alpha \beta}, (\sigma^1)_{\alpha \beta}\right)\,, \quad
(\gamma^\mu)_\a^{\ph\a\b} 
= (\gamma^\mu)_{\a\c} \varepsilon^{\c\b}\,, \quad
(\gamma^\mu)^{\a\b} 
= \varepsilon^{\a\c} (\gamma^\mu)_{\c}^{\ph\c\b} \,.
\end{equation}
With these conventions the gamma matrices are real and symmetric.
Here and in what follows we are raising and lowering spinor indices as $\theta^\a = \varepsilon^{\a\b}\theta_\b$ and $\theta_\a = \varepsilon_{\a\b}\theta^\b$, where
\begin{equation}
 \varepsilon^{12} = 1, \quad \varepsilon_{12} = -1 \,.
\end{equation}
The contraction of two spinors is defined as $\theta^2 = \varepsilon_{\a\b} \theta^\a \theta^\b$.
Finally, the spacetime Levi-Civita tensor is defined by:
\begin{equation}
\varepsilon_{012} = -1, \quad \varepsilon^{012} = 1\, .
\end{equation}

\subsection{Superconformal algebra}
\label{sec:app-algebra}

The three dimensional Lorenz group $SO(2,1)$ is generated by $\mathcal{M}_{\mu \nu}$.
A generic element of the algebra $\Jm$ contains vector indices $\mu,\nu,\lambda,\ldots$ and spinor indices $\a,\b,\ldots$, and each of them transforms under rotations as:
\begin{align}
\begin{split}
 	& [\mathcal{M}_{\mu \nu} , \Jm_{\lambda\ldots}] 
	=  i \left(
	    \eta_{\mu \lambda} \Jm_{\nu\ldots} 
	  - \eta_{\nu \lambda} \Jm_{\mu\ldots} 
    \right)\,, \\
	& [\mathcal{M}_{\mu \nu}, \Jm_{\alpha\ldots}] 
	=  +\frac{i}{2} \varepsilon_{\mu \nu \lambda} 
	     (\gamma^{\lambda})_{\alpha}^{\ph\a \beta}  \Jm_{\beta\ldots}, \\
    & [\mathcal{M}_{\mu \nu}, \Jm^{\alpha\ldots}] 
	=  -  \frac{i}{2} \varepsilon_{\mu \nu \lambda} 
	     (\gamma^{\lambda})_{\b}^{\ph\b \a}  \Jm^{\beta\ldots}\,.
\end{split}
\end{align}
With these identities it is easy to write down any commutator involving $\Mm_{\mu\nu}$.
The rest of the $3d$ conformal algebra is:
\begin{equation}\label{eq:confalg}
[\mathcal{D}, \mathcal{P}_\mu] = i \mathcal{P}_\mu, \quad [\Dop,\Kop_\mu] = - i \Kop_\mu, \quad  [\Kop_\mu, \Pop_\nu] = - 2 i \left(\Mop_{\mu \nu} + \eta_{\mu \nu} \Dop\right)\,.
\end{equation}
The $3d$ $\Nm = 2$ superconformal algebra is given by $OSP(2|4)$.
Besides the conformal and $R$-symmetry generators, it contains four Poincaré supercharges and four superconformal supercharges that anticommute as
\begin{align}
\begin{split}
 	& \{\mathcal{Q}_{\alpha},\bar{\mathcal{Q}}_{\beta}\} = 2 (\gamma^{\mu})_{\alpha \beta} \mathcal{P}_{\mu}, \quad \{\mathcal{S}^{\alpha}, \bar{\mathcal{S}}^{\beta}\} = 2  (\gamma^{\mu})^{\alpha \beta} \mathcal{K}_{\mu},  \\
	&\{ \Qop_{\alpha},\Sbop^{\beta} \} = - i \left(
        2 \delta_{\alpha}^{\ph\a\beta} (\Dop + i \Rop) 
        - \varepsilon^{\mu \nu \lambda}  (\gamma_{\lambda})_{\alpha}^{\ph\a\beta}  \Mop_{\mu \nu}
    \right)\,, \\
	&\{ \Qbop_{\alpha},\Sop^{\beta} \} = - i \left(
        2  \delta_{\alpha}^{\ph\a\beta} (\Dop - i \Rop)
        - \varepsilon^{\mu \nu \lambda} (\gamma_{\lambda})_{\alpha}^{\ph\a \beta} \Mop_{\mu \nu}
    \right)\,.
\end{split}
\end{align}
The commutation relations between the conformal group and the supercharges are
\begin{align}
	&[\Dop,\Qop_{\alpha}] = \frac{1}{2} i \Qop_{\alpha}, \quad [\Dop,\Sop^{\alpha}] = - \half i \Sop^{\alpha}, \quad [\Kop_{\mu},\Qop_{\alpha}] =  (\gamma_{\mu})_{\alpha \beta} \Sop^\beta, \quad [\Pop_{\mu},\Sop^{\alpha}] = - (\gamma_{\mu})^{\alpha \beta} \Qop_{\beta}, \nonumber \\
	&[\Dop,\Qbop_{\alpha}] = \frac{1}{2} i \Qbop_{\alpha}, \quad [\Dop,\Sbop^{\alpha}] = - \half i \Sbop^{\alpha}, \quad [\Kop_{\mu},\Qbop_{\alpha}] =  (\gamma_{\mu})_{\alpha \beta} \Sbop^\beta, \quad [\Pop_{\mu},\Sbop^{\alpha}] = - (\gamma_{\mu})^{\alpha \beta} \Qbop_{\beta}\,.
\end{align}
Lastly, all generators are neutral under $R$-symmetry, except the eight supercharges:
\begin{equation}
[\mathcal{R},\mathcal{Q}_{\alpha}] = - \mathcal{Q}_{\alpha}\,, \quad [\mathcal{R},\bar{\mathcal{Q}}_{\alpha}] =  \bar{\mathcal{Q}}_{\alpha}\,, \quad [\mathcal{R}, \mathcal{S}^{\alpha}] = - \mathcal{S}^{\alpha}\,, \quad [\Rop, \bar{\mathcal{S}}^{\alpha}] =  \bar{\mathcal{S}}^{\alpha}\,.
\end{equation}

\subsection{Differential operators}
\label{sec:diff-ops}

In this appendix we present the action of our generators in terms of differential operators in superspace.
We consider an operator $\Om(z)$ of dimension $\Delta$ and charge $r$ that transforms under rotations in a representation dictated by matrices $s_{\mu\nu}$, which satisfy the same commutation relations as $\Mm_{\mu\nu}$.
Then, the generators of the algebra act as:
\begin{align}
\label{eq:diff-ops}
 \left[\Dop,\Oop(z)\right] &= i \left(\Delta + x^\mu \partial_\mu + \frac{1}{2} \theta^\a \partial_\a + \frac{1}{2} \bar{\theta}^\a \bar{\partial}_\a \right) \Oop(z)\,,\\
\left[\Kop_{\mu},\Oop(z)\right] &= \Big(-2 i \Delta  x_{\mu}  - 2 i  x_{\mu} x^\nu \partial_\nu + i x^2 \partial_{\mu} \nonumber \\
&- i x_{\mu} ( \theta^\a \partial_\a +  \bar{\theta}^\a \bar{\partial}_\a ) 
+ i  \epsilon_{\mu \nu \rho} (\gamma^{\nu})_\alpha^{\ph \a \beta} x^{\rho} (\theta^{\alpha} \partial_{\beta} + \bar{\theta}^{\alpha} \bar{\partial}_{\beta}) \nonumber \\
&+  (\gamma_{\mu})_{\alpha \beta} \theta^{\alpha} \bar{\theta}^{\beta} ( \theta^\c \partial_\c - \bar{\theta}^\c \bar{\partial}_\c)  - \frac{i}{2}  \theta^{2}  \bar{\theta}^{2} \partial_{\mu}\nonumber\\
&- 2 \mathit{r}  (\gamma_{\mu})_{\alpha \beta} \theta^{\alpha} \bar{\theta}^{\beta}  - 2 x^{\nu} \mathit{s}_{\mu \nu} - i \eta_{\mu \nu} \varepsilon^{\nu \rho \sigma} \mathit{s}_{\rho \sigma} \theta^\a \bar\theta_\a  \Big) \Oop(z)\,,  \\
\left[\Mop_{\mu \nu} ,\Oop(z)\right] &= \left(\frac{i}{2}   \varepsilon_{\mu \nu \rho} (\gamma^{\rho})_{\alpha}^{\ph\a\beta}  (\theta^{\alpha} \partial_{\beta} + \bar{\theta}^{\alpha} \bar{\partial}_{\beta})  + i x_{\mu} \partial_{\nu} - i   x_{\nu} \partial_{\mu} + \mathit{s}_{\mu \nu}\right)\Oop(z)\,,\\
\left[\Rop,\Oop(z)\right] &=  \left(\bar{\theta}^\a \bar{\partial}_\a - \theta^\a \partial_\a + \mathit{r}\right) \Oop(z)\,,\\
\left[\Sop^{\alpha},\Oop(z)\right] &= \Big( -i   (\gamma_{\mu})^{\alpha \beta} x^{\mu} \partial_{\beta} +  \bar{\theta}^{\alpha} x^\mu \partial_\mu + \frac{1}{2} (\gamma^{\mu})^{\alpha \beta}  (\gamma^{\nu})_{\beta \gamma} \bar{\theta}^{\gamma} (x_{\mu} \partial_{\nu} - x_{\nu} \partial_{\mu}) \nonumber \\
&- \theta^{\alpha} \bar{\theta}^\b \partial_\b + \bar{\theta}^{\alpha} \theta^\b \partial_\b - 2 \bar{\theta}^{\alpha} \bar{\theta}^\b \bar{\partial}_\b \nonumber \\
&- i (\gamma^{\mu})_{\beta \gamma} \theta^{\beta} \bar{\theta}^{\gamma} \bar{\theta}^{\alpha} \partial_{\mu} + 2 \Delta  \bar{\theta}^{\alpha} - 2 \mathit{r} \bar{\theta}^{\alpha} + i \varepsilon^{\mu \nu \rho}  (\gamma_{\rho})_{\beta}^{\ph\b\a} \bar{\theta}^{\beta} \mathit{s}_{\mu \nu} \Big) \Oop(z)\,, \\
\left[\Sbop^{\alpha},\Oop(z)\right] &= \Big( i  (\gamma^{\mu})^{\alpha \beta}  x_{\mu} \bar{\partial}_{\beta} -  \theta^{\alpha} x^\mu \partial_\mu - \frac{1}{2}    (\gamma^{\mu})^{\alpha \beta}  (\gamma^{\nu})_{\beta \gamma} \theta^{\gamma} (x_{\mu} \partial_{\nu} - x_\nu \partial_\mu) \nonumber \\
& + \bar{\theta}^{\alpha} \theta^\b \bar{\partial}_\b - \theta^{\alpha} \bar{\theta}^\b \bar{\partial}_\b + 2 \theta^{\alpha} \theta^\b \partial_\b \nonumber\\
&+ i (\gamma^{\mu})_{\beta \gamma} \bar{\theta}^{\beta} \theta^{\gamma} \theta^{\alpha} \partial_{\mu} - 2 \Delta  \theta^{\alpha} - 2 \mathit{r} \theta^{\alpha} - i \varepsilon^{\mu \nu \rho}   (\gamma_{\rho})_\beta^{\ph\b\a} \theta^{\beta} \mathit{s}_{\mu \nu} \Big)\Oop(z)\,.
\end{align}

\section{Non-supersymmetric conformal blocks}
\label{sec:bosblocks}

In this appendix, we will derive the non-supersymmetric bulk and boundary blocks for bulk two-point functions and bulk-boundary-boundary correlators.

\subsection{Two-point function}

The conformal blocks for a two-point function were first derived in~\cite{McAvity:1995zd} but we will follow~\cite{Liendo:2012hy} in our approach. 

\paragraph{Bulk channel:} The bulk-channel blocks can be found by acting on the two-point function~\eqref{eq:two-pt-bos} with the bulk Casimir operator:
\begin{equation}\label{eq:bulkcasbos}
 \mathcal{C}_{\text{bulk,bos}}^{(12)} 
 = 
 -\mathcal{D}^2 
 - \half \left\{ \Kop^{\mu} , \Pop_{\mu} \right\}
 + \half \mathcal{M}^{\mu \nu} \mathcal{M}_{\mu \nu}\,.
\end{equation}
The differential operators are well know, but they can also be obtained from section~\ref{sec:diff-ops} by setting all Grassmann coordinates to zero.
The Casimir eigenvalues is $C_{\Delta,\ell} = \Delta (\Delta - d) + \ell (\ell + d -2)$, but only operators with $\ell = 0$ can appear in the bulk OPE.
For the bulk channel, it is convenient to define the blocks in terms of $g^{\Delta_{12}}_\Delta(\xi) = \xi^{(\Delta_1+\Delta_2)/2} f^{\Delta_{12}}_\Delta(\xi)$.
The resulting differential equation is
\begin{align}
\begin{split}
 \frac{
 \left( \Cm_{\text{bulk,bos}} - C_{\Delta,\ell} \right) \langle \phi_1(x_1) \phi_2(x_2)\rangle
 }{{(2x_1^\bot)^{-\Delta_1} (2 x_2^\bot)^{-\Delta_2} \xi^{-(\Delta_1+\Delta_2)/2}}}
 & = \bigg[ 4 \xi ^2 (\xi +1) \partial_\xi^2
   + 2\xi ( 2 \xi  - d + 2)  \partial_\xi \\
 & \qquad \quad 
   -  \big(\Delta(\Delta -d) + \Delta_{12}^2 \xi \big) \bigg] g^{\Delta_{12}}_\Delta(\xi )
   = 0\,,
\end{split}
\end{align}
which is solved by
\begin{align}
\label{eq:bosbulkblocks}
 g^{\Delta_{12}}_\Delta(\xi)
 = \xi ^{\Delta /2} \, _2F_1\Big(
    \frac{\Delta+\Delta_{12}}{2},
    \frac{\Delta-\Delta_{12}}{2};
    \Delta + 1 - \frac{d}{2};
    -\xi 
 \Big).
\end{align}
Whenever the superscript $\Delta_{12}$ is omitted, it is assumed that $\Delta_{12} = 0$.
\paragraph{Boundary channel:} In the boundary channel, the conformal blocks are eigenfunctions of the boundary Casimir that acts on a single point
\begin{equation}\label{eq:bdycasbos}
 \hat{\mathcal{C}}^{(1)}_{\text{non-susy}} 
 = 
 - \mathcal{D}^2 
 - \half \left\{ \mathcal{K}^{a} ,\mathcal{P}_{a} \right\}  
 + \half \mathcal{M}^{a b} \mathcal{M}_{ab }\,,
\end{equation}
where the index $a,b$ runs only on directions parallel to the boundary.
The eigenvalue of the boundary Casimir is $\hat C_{\Delta,j} = \Delta (\Delta - d + 1) + j(j + d - 1)$, but once more we have to take $j = 0$ because only scalar operators appear in the BOE of a bulk scalar.
The resulting differential equation is
\begin{align}
\begin{split}
 \frac{
 \left( \hat \Cm_{\text{non-susy}} - \hat C_{\Delta,0} \right) \langle \phi_1(x_1) \phi_2(x_2)\rangle
 }{(2x_1^\bot)^{-\Delta_1} (2 x_2^\bot)^{-\Delta_2}}
 & = \bigg[ \xi  (\xi +1) \partial_\xi^2
 + \frac{d}{2} (2 \xi +1) \partial_\xi \\
 & \qquad \qquad \qquad \quad
 - \hat\Delta (\hat\Delta - d + 1) \bigg] \hat f_{\hat \Delta}(\xi )
 = 0\,,
\end{split}
\end{align}
which is solved by
\begin{align}
 \hat f_{\hat\Delta}(\xi) =
 \xi ^{-\hat\Delta } \, 
 _2F_1\Big(\hat\Delta ,\hat\Delta-\frac{d}{2} + 1;2 \hat\Delta -d + 2; -\frac{1}{\xi }\Big)\,.
\end{align}

\subsection{Three-point bosonic blocks}
\label{app:3pt_bos_blocks}
 
In this section we restrict to $d = 3$. We start with considering the bosonic correlator
\begin{equation}
\langle \Oop_1 (x) \hat{\Oop}_{2,j} (0) \hat{\Oop}_3 (\infty) \rangle =  \frac{\left(x^a \omega_a \right)^j}{(x^{\bot})^{\Delta_1 + \hat{\Delta}_{23}}|x^a|^j} \mathcal{F}^{\text{3pt}} (\chi), \qquad
\chi =   \frac{|x^a|^2}{(x^\bot)^2}\,,
\end{equation}
where the second operator has parallel spin $j$. We used index-free notation to contract all vector indices, and $\omega_a$ is a null-vector. 
We need to evaluate the eigenvalue equation
\begin{equation}
\hat{\mathcal{C}}_{\text{bos}}^{(1)} \langle \Oop_1 (x) \hat{\Oop}_{2,j} (0) \hat{\Oop}_3 (\infty) \rangle = \hat{C}_{\hat{\Delta},0}\langle \Oop_1 (x) \hat{\Oop}_{2,j} (0) \hat{\Oop}_3 (\infty) \rangle\,,
\end{equation}
where $\hat{C}_{\hat{\Delta},0}$ is the boundary Casimir eigenvalue when the parallel spin of the exchanged operator is zero. This gives us the differential equation
\begin{equation}\label{eq:diffeqbos3pt02}
\left[4 \chi  (\chi +1) \partial^{2}_{\chi} + 4 \left((\hat{\Delta}_{23}+2) \chi +1\right) \partial_{\chi} -   \hat{\Delta}(\hat{\Delta}-2) + \hat{\Delta}_{23} (\hat{\Delta}_{23}+2) - \frac{j^2}{\chi}\right] \hat{f}^{\text{3pt},\hat{\Delta}_{23}}_{\hat{\Delta},j} = 0\,.
\end{equation}
The solution to equation \eqref{eq:diffeqbos3pt02} is once more given by a hypergeometric function
\begin{equation}\label{eq:threeptbosblocks}
\hat{f}^{\text{3pt},\hat{\Delta}_{23}}_{\hat{\Delta},j} (\chi) = \chi^{-\half (\hat{\Delta} + \hat{\Delta}_{23})} \, _2F_1 \Big(\half \left(\hat{\Delta} + \hat{\Delta}_{23} - j \right), \half \left(\hat{\Delta} + \hat{\Delta}_{23} + j \right);\hat{\Delta}; - \frac{1}{\chi} \Big)\,.
\end{equation}

\section{More on blocks across dimensions}
\label{sec:app-blocks-4-eps}

In this appendix, we provide more details on the derivation of the superconformal blocks in any number of dimension.
In section~\ref{sec:4-eps} we showed that the supersymmetric part of the Casimir acting on a two-point function can be written in terms of $Q$ supercharges acting on the two-point function.
Our current goal is to find equivalent expressions where the supercharges are replaced by a differential operator, for example
\begin{align}
\label{eq:QsActing}
\begin{split}
 \big\langle 0 \big|
    \left[ Q^-_\ad, \phi_1(x_1) \right]
    \left[ Q^+_\a, \bar \phi_2(x_2) \right]
 \big| 0 \big\rangle
 \sim \Dm_{x}
 \big\langle 0 \big| \phi_1(x_1) \bar \phi_2(x_2) \big| 0 \big\rangle \,. 
\end{split}
\end{align}
It was proposed in~\cite{Bobev:2015jxa} that this can be achieved with supersymmetric Ward identities. Here we give a quick summary of the strategy.
In our setup the supercharges $\Qm_A$ and $\Sm_A$ are preserved by the boundary, so the following Ward identities are satisfied:
\begin{align}
\begin{split}
 & \big\langle 0 \big| 
    \big\{ Q^{\text{bdy}}_1, \left[ Q^-_1, \phi_1(x_1) \right] \bar \phi_2(x_2) \big\} 
  \big| 0 \big\rangle = 0 \,, \\
& \big\langle 0 \big| 
    \big\{  Q^{\text{bdy}}_1, \left[ Q^-_2, \phi_1(x_1) \right] \bar \phi_2(x_2) \big\} 
  \big| 0 \big\rangle = 0\,, \\
& \big\langle 0 \big| 
    \big\{  Q^{\text{bdy}}_2, \left[ Q^-_1, \phi_1(x_1) \right] \bar \phi_2(x_2) \big\} 
  \big| 0 \big\rangle = 0\,, \\
\end{split}
\begin{split}
 & \big\langle 0 \big| 
    \big\{  S^{\text{bdy}}_1, \left[ Q^-_1, \phi_1(x_1) \right] \bar \phi_2(x_2) \big\} 
  \big| 0 \big\rangle = 0\,, \\
& \big\langle 0 \big| 
    \big\{ S^{\text{bdy}}_1, \left[ Q^-_2, \phi_1(x_1) \right] \bar \phi_2(x_2) \big\} 
  \big| 0 \big\rangle = 0\,. \\
  \ph{1}
\end{split}
\end{align}
There are other Ward identities that can be considered, but these five are sufficient for our purposes.
At this point, it is hard to continue without an explicit matrix representation for the Clifford algebra, so we focus on $d=3$ where $\Sigma_\mu = (\sigma_3, \sigma_1, \sigma_2)$.
Let us consider explicitly the simplest Ward identity to show how to replace the supercharges with differential operators in the general case.
With elementary manipulations we find:
\begin{align}
\begin{split}
   0 
& = \big\langle 0 \big| 
    \big\{ Q^{\text{bdy}}_2, \left[ Q^-_1, \phi_1(x_1) \right] \bar \phi_2(x_2) \big\} 
    \big| 0 \big\rangle \\
& = \big\langle 0 \big| 
    \big\{  Q^{\text{bdy}}_2, \left[ Q^-_1, \phi_1(x_1) \right] \big\} \bar \phi_2(x_2)
    \big| 0 \big\rangle
  - \big\langle 0 \big| 
    \big[ Q^-_1, \phi_1(x_1) \big] \big[  Q^{\text{bdy}}_2, \bar \phi_2(x_2) \big]
    \big| 0 \big\rangle \\
& = \big\langle 0 \big| 
    \big[ \big\{ Q_2^+, Q^-_1 \big\}, \phi_1(x_1) \big] \bar \phi_2(x_2)
    \big| 0 \big\rangle
  - \big\langle 0 \big| 
    \big[ Q^-_1, \phi_1(x_1) \big] \big[ Q_2^+, \bar \phi_2(x_2) \big]
    \big| 0 \big\rangle\,. 
\end{split}
\end{align}
In our conventions $\{ Q_2^+, Q^-_1 \} = P_2 + i P_3$ and also $[P_\mu, \Om(x)] = -i \partial_\mu \Om(x)$, so we conclude
\begin{align}
 \left\langle 0 \left| 
    \left[ Q^-_1, \phi_1(x_1) \right] \left[ Q_2^+, \bar \phi_2(x_2) \right]
 \right| 0 \right\rangle
 = \left( -i \partial_2 + \partial_3 \right)   
 \left\langle 0 \left| \phi_1(x_1) \bar \phi_2(x_2) \right| 0 \right\rangle\,.
\end{align}
The other Ward identities can be manipulated identically, but unlike the example we showed they do not decouple, so one has to solve a simple linear system of equations to obtain the terms we are interested in.

These steps can be automated in \texttt{Mathematica} and applied to all cases of interest in $d=3,4$.
The resulting differential operators $\Dm_{x}$ depend on $x_i^\mu$ and $\partial_{\mu,i}$ and take a complicated looking form.
However, we know that the Casimir operator $C_{\text{bulk/bdy},\text{susy}}$ has to respect conformal invariance, so when we combine all the contributing terms, the result has to be a differential operator of the cross-ratio $\xi$.
Indeed, in $d=3,4$ we find the following results:
\begin{align}
\begin{split}
 i x_{12}^\mu \bar\Sigma_\mu^{\ad\a}
 \left\langle 0 \left|
    \left[ Q^-_\ad, \phi_1(x_1) \right]
    \left[ Q^+_\a, \bar \phi_2(x_2) \right]
 \right| 0 \right\rangle
&\to \big( 4 \xi  \partial_\xi - 2 (\Delta_1 + \Delta_2) \big) \Gm(\xi )\,, \\
 i x_1^\bot \left\langle 0 \left|
    \left\{ Q^-_1, \left[ Q^-_2, \phi_1(x_1) \right] \right\} \bar \phi_2(x_2) 
 \right| 0 \right\rangle
&\to (- \xi  \partial_\xi - \Delta_1) \Fm(\xi) , \\
 i x_1^\bot \left\langle 0 \left|
    \left\{ Q^-_1, \left[ Q^-_2, \phi_1(x_1) \right] \right\} \phi_2(x_2) 
 \right| 0 \right\rangle
&\to \big({-}(\xi+1)  \partial_\xi - \Delta_1 \big) \Fm(\xi)\,.
\end{split}
\end{align}
From these results we can obtain~\eqref{eq:bdy-contrib-PPb},~\eqref{eq:bdy-contrib-PP} and~\eqref{eq:blk-contrib-PPb}.
Although the intermediate differential operators were complicated, the final result takes a remarkably simple form.
Perhaps one could find a more direct method of obtaining these results, and at the same time make it more manifest that the result is indeed independent of $d$.


\providecommand{\href}[2]{#2}\begingroup\raggedright\endgroup

\end{document}